\documentclass{article}%
\usepackage{amssymb}
\usepackage{amsmath}
\usepackage{amsfonts}
\usepackage{graphicx}%
\providecommand{\U}[1]{\protect\rule{.1in}{.1in}}

\begin{document}

\title{A multiple scattering theory approach to solving the \\time-dependent Schr\"{o}dinger equation with an asymmetric\\rectangular potential}
\author{Victor F. Los (1) and Nicholas\ V. Los (2)
\and ((1) Institute of Magnetism, Nat. Acad. Sci. Ukraine, Kiev, Ukraine,
\and (2) Luxoft Eastern Europe, Kiev, Ukraine)}

\begin{abstract}
An exact time-dependent solution for the wave function $\psi(\mathbf{r},t)$ of
a \ particle moving in the presence of an asymmetric rectangular well/barrier
potential varying in one dimension is obtained by applying a novel for this
problem approach using multiple scattering theory (MST) for the calculation of
the space-time propagator. This approach, based on the localized at the
potential jumps effective potentials responsible for transmission through and
reflection from the considered rectangular potential, enables considering
these processes from a particle (rather than a wave) point of view. The
solution describes these quantum phenomena as a function of time and is
related to the fundamental issues (such as measuring time) of quantum
mechanics. It is presented in terms of integrals of elementary functions and
is a sum of the forward- and backward-moving components of the wave packet.
The relative contribution of these components and their interference as well
as of the potential asymmetry to the probability density $\left\vert
\psi(x,t)\right\vert ^{2}$ and particle dwell time is considered and
numerically visualized for narrow and broad energy (momentum) distributions of
the initial Gaussian wave packet. The obtained solution is also related to the
kinetic theory of nanostructures due to the fact that the considered potential
can model the spin-dependent potential profile of the magnetic multilayers
used in spintronics devices.

Keywords:

multiple-scattering theory, time-dependent Schrodinger equation, \ 

rectangular asymmetric well/barrier potential; backward-moving \ wave,

dwell time, magnetic nanostructures

Corresponding author

Victor F. Los, victorlos@mail.ru

Phone/Fax:+38-044-424-1020

\end{abstract}
\maketitle

\section{Introduction}

The time-dependent aspects of reflection from and transmission through a
potential step/barrier/well raised several questions that have not yet been
completely clarified and have recently acquired relevance in view of renewed
interest in the fundamental problem of measuring time in quantum mechanics
(see \cite{Time in QM}). The tunneling time can serve as an example (see, e.g.
a review \cite{Hauge 1989}).

The mentioned phenomena are less surprising when we think of a wave being,
e.g., reflected from a downward potential step. In the stationary case, these
quantum phenomena easily follow from standard textbook analysis, which reduces
to solving the stationary Schr\"{o}dinger equation by matching the wave
function of a plane wave of energy $E$ and its derivative across the potential
jumps. However, in this case, there are no real transport phenomena, i.e. in
the absence of energy dispersion, $\Delta E=0$, the particle time of
transmission through or of arrival (TOA) to the potential jumps is indefinite
($\Delta t\backsim\hbar/\Delta E$).

These processes are more surprising from the particle point of view, and it is
interesting to verify the mentioned non-classical phenomena by considering the
time-dependent picture in a realistic situation, when a particle, originally
localized outside the potential well/barrier, moves towards the potential and
experiences scattering at the potential jumps. In order to describe these
time-dependent processes, the corresponding time-dependent Schr\"{o}dinger
equation with a rectangular potential should be solved, which is much more
involved compared to the conventional stationary case. From the particle point
of view, it also seems desirable to be able to apply the multiple scattering
theory (MST) to this principally important situation. The MST is
conventionally formulated in terms of a "free" particle Green function
(propagator) modified by the scatterings at the inhomogeneities such as the
interfaces between different media. It is clear that the scattering events
occur in the interface area and are stipulated by some fairly localized
potentials. Thus one faces an interesting problem of finding a localized
potential responsible for particle scattering from a potential inhomogeneity.

In addition, there is one striking and classically forbidden counterintuitive
(and often overlooked) effect even in the simplest 1D time-dependent
scattering by the mentioned potentials. A wave packet representing an ensemble
of particles, confined initially (at $t=t_{0}$), say, somewhere to the region
$x<0$, consists of both positive and negative momentum components due to the
fact that a particle cannot be completely localized at $x<0$ if the wave
packet contains only $p>0$ components. One would expect that only particles
with positive momenta $p$ may arrive at positive positions $x>0$ at $t>t_{0}$.
However, the wave packet's negative momentum components (restricted to a half
line in momentum space) are necessarily different from zero in the whole $x$
space ($-\infty\div\infty$), represent the particles' presence at $x>0$ at
initial moment of time $t_{0}$, and, therefore, may contribute, for example,
to the distribution of the particles' time of arrival (TOA) to $x>0$
\cite{Muga Sources 2001,Baute 2002}. It is worth noting that the contribution
of the backward-moving (negative momentum) components in the initial-value
problem is in some sense equivalent to the contribution of the negative energy
(evanescent) components in the source solution \cite{Muga Sources 2001}. Thus,
the correct treatment of some aspects of the kinetics of the wave packet (even
in the 1D case and even for "free" motion) becomes a nontrivial problem and is
closely related to the fundamental problem of measuring time in quantum
mechanics, such as TOA, the dwell time, and tunneling time.

On the other hand, the mentioned rather academic (but fundamentally important)
problems have acquired reality and significance due to important practical
applications in the newly emerged field of nanoscience and nanotechnology.
Rectangular potential barriers/wells may often satisfactorily approximate the
one-dimensional potential profiles in layered magnetic nanostructures (with
sharp interfaces). In such nanostructures, the giant magnetoresistance (GMR)
\cite{GMR} and tunneling magnetoresistance (TMR) \cite{TMR} effects occur.
These effects, which stem, particularly, from quantum mechanical
spin-dependent electrons tunneling through potential barriers or their
reflection from potential wells, have led to very important commercial
applications of spintronic devices.

The solution to the time-dependent Schr\"{o}dinger equation can be obtained
with the help of the spacetime propagator (Green's function), which has been
conveniently calculated by the path-integral method. The list of exact
solutions for this propagator is very short. For example, there is an exact
solution to the space-time propagator by the path-integral method in the
one-dimensional square barrier case obtained in \cite{Barut 1988}, but this
solution is very complicated, implicit and not easy to analyze (see also
\cite{Schulman 1982, Carvalho 1993, Yearsley 2008}).

Recently, we have suggested a method \cite{Los 2010} for the calculation of
the spacetime propagator which is based on the energy integration of the
spectral density matrix (discontinuity of the energy-dependent Green function
across the real energy axis). The energy-dependent Green function is then
easily obtained for the step/barrier/well potentials with multiple-scattering
theory (MST) using the effective energy-dependent potentials found in
\cite{Los 2010}, which are responsible for reflection from and transmission
through the potential step. The obtained $\delta$-like potentials describing
the quantum scattering from a potential step provide a clear picture of the
particle's scattering taking place at the interfaces and make it convenient to
calculate the energy-dependent Green function and the space-time propagator
especially when the scattering from more than one interface needs to be
accounted for and there are other sources of scattering by the point-like
scatterers. Such a situation is typical for real multilayers with disordered
interfaces \cite{Butler and Los (2003), Los (2005), Los (2008)} and for the
Casimir effect \cite{Casimir}. An important advantage of our approach to
propagator calculation is also that it allows for a natural decomposition of
the general initial wave function evolution in time into both the forward-
($p=\hbar k>0$) and backward-moving (negative momentum $p<0$) components and
an analysis of the contribution of both these terms and their interference to
particle reflection and transmission. This approach has been further applied
to the analysis of the time-dependent properties of the scattering by the
imaginary step \cite{Los 2011} (related to calculation of the particle time of
arrival) and rectangular symmetric barrier/well potentials \cite{Los 2012,Los
2013}.

In this paper, we generalize our approach to the solution of the
time-dependent Schr\"{o}dinger equation in the case when a particle moves
towards a rectangular asymmetric (spin-dependent) potential. Tunneling through
asymmetric potentials has already important applications in semiconductor
heterostructures. The asymmetric (spin-dependent) rectangular potential
barrier/well can also model the potential profile of the magnetic threelayer
switched from the parallel configuration of magnetic layers (symmetric
potential) to the anti-parallel configuration of layers. Although this
potential, which models the spin-dependent potential profile in magnetic
nanostructures, changes only in the $x$ (perpendicular to interfaces)
direction, the system under consideration in this paper is a real
three-dimensional one. A simple exact solution for the time-dependent
propagator in terms of integrals of elementary functions is obtained, which is
valid for both the well and barrier cases. This solution fully resolves the
corresponding time-dependent Schr\"{o}dinger equation and provides exact
analytical expressions for the wave function $\psi(\mathbf{r},t)$ in the
spatial regions before, inside and after the potential with account for the
backward-moving terms caused by the negative momentum components of the
initial wave function. It is important that the obtained solution allows for
numerical visualization of the observables defined by the wave function
$\psi(\mathbf{r},t)$ in the mentioned spatial regions. Thus, the corresponding
probability densities $\left\vert \psi(\mathbf{r},t)\right\vert ^{2}$ are
analyzed and numerically visualized for the Gaussian initial wave packet with
special attention to the counterintuitive contribution (see \cite{Muga Sources
2001,Baute 2002}) of the backward-moving wave packet components and the
potential asymmetry. We show (and visualize) that the contribution of the
backward-moving components of the wave packet is small in the quasiclassical
case but is otherwise important. It is also shown that the influence of the
potential asymmetry is more pronounced when the contribution of the
backward-moving wave packet components is essential. The dwell time, which
characterizes the average time spent by a particle in the potential region and
is related to the enduring quantum physics problem of calculating the
tunneling time, is considered for the asymmetric rectangular potential. The
obtained results can also provide a foundation for a kinetic theory of nanostructures.

\section{Multiple-scattering calculation of the space-time propagator and
time-dependent solution for the Schr\"{o}dinger equation}

We consider a particle moving toward the following asymmetric one-dimensional
rectangular potential of the width $d$ placed in the interval $(0<x<d)$%
\begin{equation}
V(x)=[\theta(x)-\theta(x-d)]U+\theta(x-d)\Delta, \label{1}%
\end{equation}
where $\theta(x)$ is the Heaviside step function, and the potential parameter
$U$ can acquire positive (barrier) as well as negative (well) values. The wave
packet, modeling a particle, will approach a potential (\ref{1}) from the left
(where the particle potential energy is zero) and the parameter $\Delta$ is
supposed to be non-negative ($\Delta\geq0$). With the potential (\ref{1}) we
can model, e.g, the spin-dependent potential of a threelayer, which consists
of a spacer (metallic or insulator) sandwiched between two magnetic (infinite)
layers. An asymmetry (spin-dependence) of the potential (\ref{1}) is defined
by the parameter $\Delta$ via the electron spectrum in different magnetic
layers as
\begin{align}
k_{<}^{0}(E;\mathbf{k}_{||})  &  =k(E;\mathbf{k}_{||}),k_{>}^{d}%
(E;\mathbf{k}_{||})=k_{\Delta}(E;\mathbf{k}_{||}),\nonumber\\
k(E;\mathbf{k}_{||})  &  =\sqrt{\frac{2m}{\hbar^{2}}E-\mathbf{k}_{||}^{2}%
},k_{\Delta}(E;\mathbf{k}_{||})=\sqrt{\frac{2m}{\hbar^{2}}(E-\Delta
)-\mathbf{k}_{||}^{2}},\nonumber\\
k_{>}^{0}(E;\mathbf{k}_{||})  &  =k_{<}^{d}(E;\mathbf{k}_{||})=k_{u}%
(E;\mathbf{k}_{||}),k_{u}(E;\mathbf{k}_{||})=\sqrt{\frac{2m}{\hbar^{2}%
}(E-U)-\mathbf{k}_{||}^{2}}, \label{2}%
\end{align}
where $k_{>(<)}^{0}(E;\mathbf{k}_{||})$ and $k_{>(<)}^{d}(E;\mathbf{k}_{||})$
are the perpendicular-to-interfaces (located at $x=0$ and $x=d$) components of
the particle wave vector $\mathbf{k}$ to the right ($>$) or to the left ($<$)
of the corresponding interface, while $\mathbf{k}_{||}$ is the
parallel-to-interfaces component of an electron wave vector, which is
conserved for the sharp interfaces under consideration. The two-dimensional
vector $\mathbf{k}_{||}$ defines the angle of electron incidence at the interface.

From the particle propagation point of view, the partial reflection from and
transmission through a potential inhomogeneity may be explained by the quantum
mechanical rules of computing the probabilities of different events. These
rules represent the quantum mechanical generalization of the Huygens-Fresnel
principle and were introduced by Feynman as the path-integral formalism
\cite{Feynman}. It states that a wave function of a single particle moving in
a perturbing potential $V(\mathbf{r},t)$ may be presented as%
\begin{equation}
\Psi(\mathbf{r},t)=\int d\mathbf{r}^{\prime}K(\mathbf{r},t;\mathbf{r}^{\prime
},t_{0})\Psi(\mathbf{r}^{\prime},t_{0}). \label{3}%
\end{equation}
Equation (\ref{3}) shows (in accordance with the Huygens-Fresnel principle)
that the wave function $\Psi(\mathbf{r},t)$ at the spacetime point
($\mathbf{r},t$) is the sum of the contributions of all points of space where
the wave function $\Psi(\mathbf{r}^{\prime},t_{0})$ at $t=t_{0}$ is nonzero.
The propagator $K(\mathbf{r},t;\mathbf{r}^{\prime},t_{0})$ is the probability
amplitude for the particle's transition from the initial spacetime point
($\mathbf{r}^{\prime},t_{0}$) to the final point ($\mathbf{r},t$) by means of
all possible paths. It provides the complete information on the particle's
dynamics and resolves the corresponding time-dependent Schr\"{o}dinger equation.

Thus, the problem is to find the propagator $K(\mathbf{r},t;\mathbf{r}%
^{\prime},t_{0})$ for the given potential $V(\mathbf{r},t)$. In some cases,
for example when the potential is quadratic in the space variable, the kernel
$K(\mathbf{r},t;\mathbf{r}^{\prime},t_{0})$ may be calculated exactly. In the
case when the potential changes smoothly enough, a quasi-classical
approximation can be employed. It is not, however, the case for the singular
potential (\ref{1}).

According to \cite{Los 2010}, the time-dependent retarded (operator)
propagator $K(t;t^{\prime})=\theta(t-t^{\prime})\exp\left[  -\frac{i}{\hbar
}H(t-t^{\prime})\right]  $ can be calculated with the use of the following
definition%
\begin{equation}
K(t;t^{\prime})=\theta(t-t^{\prime})\frac{i}{2\pi}%
{\displaystyle\int\limits_{-\infty}^{\infty}}
e^{-\frac{i}{\hbar}E(t-t^{\prime})}\left[  G(E+i\varepsilon)-G(E-i\varepsilon
)\right]  dE,\varepsilon\rightarrow+0, \label{4}%
\end{equation}
where%
\begin{equation}
G(E)=\frac{1}{E-H} \label{5}%
\end{equation}
is the resolvent operator, $E$ stands for the energy and $H$ is the
Hamiltonian of the system under consideration. Correspondingly, $G(E\pm
i\varepsilon)=G^{\pm}(E)$ defines the retarded ($G^{+}$) or the advanced
($G^{-}$) Green function. The $E$-resolving Fourier transformation (\ref{4})
is useful for the calculation of the propagator $K(t;t^{\prime})$ when the
Green functions $G^{\pm}(E)$ may be found for each value of $E$, i.e. when the
considered processes are energy-conserved as is the case considered in this paper.

We are looking for the spacetime propagator $K(\mathbf{r},t;\mathbf{r}%
^{\prime},0)=<\mathbf{r}|K(t;0)|\mathbf{r}^{\prime}>$ , defining the
probability amplitude for a particle's transition from the initial point
($\mathbf{r}^{\prime},0$) to the final destination ($\mathbf{r,}t$) in the
presence of the potential (\ref{1}). For the considered geometry, it is
convenient to present the $r$-representation of the Green function with the
Hamiltonian $H$, $G(\mathbf{r,r}^{\prime};E)=$ $<\mathbf{r}|\frac{1}%
{E-H}|\mathbf{r}^{\prime}>$, as follows%
\begin{equation}
G(\mathbf{r},\mathbf{r}^{/};E)=\frac{1}{A}%
{\textstyle\sum\limits_{\mathbf{k}_{||}}}
e^{i\mathbf{k}_{||}(\mathbf{\rho}-\mathbf{\rho}^{\prime})}G(x,x^{\prime
};E;\mathbf{k}_{||}), \label{6}%
\end{equation}
where $\mathbf{\rho=(}y,z\mathbf{)}$ is a two-dimensional
parallel-to-interface vector and $A$ is the area of the interface. Thus, the
problem is reduced to finding the one-dimensional Green's function
$G(x,x^{\prime};E;\mathbf{k}_{||})$ dependent on the conserved particle energy
and parallel-to-interface component of the wave vector. In the following
calculation of this function we will suppress for simplicity the dependence on
the argument $\mathbf{k}_{||}$, which will be recovered at the end of calculation.

We showed in \cite{Los 2010} that the Hamiltonian corresponding to the
energy-conserving processes of scattering at potential steps can be presented
as
\begin{align}
H  &  =H_{0}+H_{i}(x;E),\nonumber\\
H_{i}(x;E)  &  =%
{\displaystyle\sum\limits_{s}}
H_{i}^{s}(E)\delta(x-x_{s}). \label{7}%
\end{align}
Here, $H_{i}(x;E)$ describes the perturbation of the "free" particle motion
(defined by $H_{0}=-\frac{\hbar^{2}}{2m}\frac{\partial^{2}}{\partial
\mathbf{r}^{2}}$) localized at the potential steps with coordinates $x_{s}$
(in the case of the potential (\ref{1}), there are two potential steps at
$x_{s}=0$ and $x_{s}=d$)%
\begin{align}
H_{i>}^{s}(E)  &  =\frac{i\hbar}{2}[v_{>}^{s}(E)-v_{<}^{s}(E)],\nonumber\\
H_{i<}^{s}(E)  &  =\frac{i\hbar}{2}[v_{<}^{s}(E)-v_{>}^{s}(E)],\nonumber\\
H_{i><}^{s}(E)  &  =\frac{2i\hbar v_{>}^{s}(E)v_{<}^{s}(E)}{[\sqrt{v_{>}%
^{s}(E)}+\sqrt{v_{<}^{s}(E)}]^{2}}, \label{8}%
\end{align}
where $H_{i>(<)}^{s}(E)$ is the reflection (from the potential step at
$x=x_{s}$, $s\in\{0,d\}$) potential amplitude, the index $>(<)$ indicates the
side on which the particle approaches the interface at $x=x_{s}$: right ($>$)
or left ($<$); $H_{i><}^{s}(E)$ is the transmission potential amplitude, and
the velocities $v_{>(<)}^{s}(E)=\hbar k_{>(<)}^{s}(E)/m,$ ($k_{>(<)}^{s}(E)$
are given by (\ref{2})). Note that the perturbation Hamiltonian $H_{i}^{s}$ 's
dependence on $\mathbf{k}_{||}$ (which is omitted for brevity) comes from Eq.
(\ref{2}).

The perturbation expansion for the retarded Green function $G^{+}(x,x^{\prime
};E)$ in the case of the rectangular potential (\ref{1}), which can be
effectively represented by the two-step effective Hamiltonian (\ref{7}), reads
for different source (given by $x^{\prime}$) and destination (determined by
$x$) areas of interest as (see also \cite{Los 2012,Los 2013})%
\begin{align}
G^{+}(x,x^{\prime};E)  &  =G_{0}^{+}(x,d;E)T^{+}(E)G_{0}^{+}(0,x^{\prime
};E),x^{\prime}<0,x>d,\nonumber\\
G^{+}(x,x^{\prime};E)  &  =G_{0}^{+}(x,0;E)T^{+}(E)G_{0}^{+}(d,x^{\prime
};E),x^{\prime}>d,x<0,\nonumber\\
G^{+}(x,x^{\prime};E)  &  =G_{0}^{+}(x,0;E)T^{\prime+}(E)G_{0}^{+}%
(0,x^{\prime};E)+G_{0}^{+}(x,d;E)R^{\prime+}(E)G_{0}^{+}(0,x^{\prime}%
;E),x^{/}<0,0<x<d,\nonumber\\
G^{+}(x,x^{\prime};E)  &  =G_{0}^{+}(x,0;E)T^{\prime+}(E)G_{0}^{+}%
(0,x^{\prime};E)+G_{0}^{+}(x,0;E)R^{\prime+}(E)G_{0}^{+}(d,x^{\prime
};E),0<x^{/}<d,x<0,\nonumber\\
G^{+}(x,x^{\prime};E)  &  =G_{0}^{+}(x,x^{\prime};E)+G_{0}^{+}(x,0;E)R^{+}%
(E)G_{0}^{+}(0,x^{\prime};E),x^{\prime}<0,x<0, \label{8a}%
\end{align}
where the transmission and reflection matrices are
\begin{align}
T^{+}(E)  &  =\frac{T_{><}^{d+}(E)G_{0}^{+}(d,0;E)T_{><}^{0+}(E)}{D^{+}%
(E)},\nonumber\\
T^{\prime+}(E)  &  =\frac{T_{><}^{0+}(E)}{D^{+}(E)},R^{\prime+}(E)=T_{<}%
^{d+}(E)G_{0}^{+}(d,0;E)T^{\prime+}(E),\nonumber\\
R^{+}(E)  &  =T_{<}^{0+}(E)+\frac{T_{><}^{0+}(E)G_{0}^{+}(0,d;E)T_{<}%
^{d+}(E)G_{0}^{+}(d,0;E)T_{><}^{0+}(E)}{D^{+}(E)}\nonumber\\
D^{+}(E)  &  =1-G_{0}^{+}(d,0;E)T_{>}^{0+}(E)G_{0}^{+}(0,d;E)T_{<}^{d+}(E).
\label{8b}%
\end{align}
The one-dimensional retarded Green function $G_{0}^{+}(x,x^{\prime};E)$
corresponding to a free particle moving in constant potential $V(x)=0$ or
$V(x)=U($or $\Delta)$ is (see, e.g. \cite{Economou})%
\begin{align}
G_{0}^{+}(x,x^{\prime};E)  &  =\frac{m}{i\hbar^{2}k(E)}\exp[ik(E)|x-x^{\prime
}|],V(x)=0,\nonumber\\
G_{0}^{+}(x,x^{\prime};E)  &  =\frac{m}{i\hbar^{2}k_{u(\Delta)}(E)}%
\exp[ik_{u(\Delta)}(E)|x-x^{\prime}|],V(x)=U(\text{or }\Delta), \label{9}%
\end{align}
where the wave numbers are determined by (\ref{2}). The scattering (at the
step located at $x=x_{s}$) t-matrices are defined by the following
perturbation expansion:%
\begin{align}
T^{s}(E)  &  =H_{i}^{s}(E)+H_{i}^{s}(E)G_{0}(x_{s},x_{s};E)H_{i}^{s}%
(E)+\ldots\nonumber\\
&  =\frac{H_{i}^{s}(E)}{1-G_{0}(x_{s},x_{s};E)H_{i}^{s}(E)}, \label{9a}%
\end{align}
where $H_{i}^{s}(E)$ and the interface Green function $G_{0}(x_{s},x_{s};E)$
are defined differently for reflection and transmission processes \cite{Los
2010}: the step-localized effective potential is given by Eq. (\ref{8}) and
the retarded Green functions at the interface for the considered reflection
and transmission processes are, correspondingly,
\begin{align}
G_{0>(<)}^{+}(x_{s},x_{s};E)  &  =1/i\hbar v_{>(<)}^{s}(E)\nonumber\\
G_{0><}^{+}(x_{s},x_{s};E)  &  =1/i\hbar\sqrt{v_{>}^{s}(E)v_{<}^{s}(E)}
\label{10}%
\end{align}
in accordance with (\ref{9}).

From (\ref{8}), (\ref{9a}) and (\ref{10}), we have for the reflection
$T_{>(<)}^{s+}(E)$ and transmission $T_{><}^{s+}(E)$ t-matrices, used in
(\ref{8b}) ($s\in\{0,d\}$), corresponding to the retarded Green function and
scattering at the interface located at $x=x_{s}\in\{0,d\}$,
\begin{align}
T_{>(<)}^{s+}(E)  &  =i\hbar v_{>(<)}^{s}r_{>(<)}^{s},\nonumber\\
T_{><}^{s+}(E)  &  =i\hbar\sqrt{v_{>}^{s}v_{<}^{s}}t^{s}, \label{11}%
\end{align}
where $r_{>(<)}^{s}(E)$ and $t^{s}(E)$ are the standard amplitudes for
reflection to the right (left) of the potential step at $x=x_{s}$ and
transmission through this step%
\begin{align}
r_{>}^{s}(E)  &  =\frac{k_{>}^{s}-k_{<}^{s}}{k_{>}^{s}+k_{<}^{s}},r_{<}%
^{s}(E)=\frac{k_{<}^{s}-k_{>}^{s}}{k_{>}^{s}+k_{<}^{s}},\nonumber\\
t^{s}(E)  &  =\frac{2\sqrt{k_{>}^{s}k_{<}^{s}}}{k_{>}^{s}+k_{<}^{s}},
\label{12}%
\end{align}
and the argument $E$ in the wave vectors is omitted for brevity.

Using Eqs. (\ref{2}),(\ref{8a}), (\ref{8b}), (\ref{9}), (\ref{11}) and
(\ref{12}), we obtain%
\begin{align}
G^{+}(x,x^{\prime};E)  &  =\frac{m}{i\hbar^{2}\sqrt{kk_{\Delta}}}%
e^{ik_{\Delta}(x-d)}t(E)e^{-ikx^{\prime}},x^{\prime}<0,x>d,\nonumber\\
G^{+}(x,x^{\prime};E)  &  =\frac{m}{i\hbar^{2}\sqrt{kk_{\Delta}}}%
e^{-ikx}t(E)e^{ik_{\Delta}(x^{\prime}-d)},x^{\prime}>d,x<0,\nonumber\\
G^{+}(x,x^{\prime};E)  &  =\frac{m}{i\hbar^{2}\sqrt{kk_{u}}}\left[
e^{ik_{u}x}t^{\prime}(E)e^{-ikx^{\prime}}+e^{-ik_{u}x}r^{\prime}%
(E)e^{-ikx^{\prime}}\right]  ,x^{\prime}<0,0<x<d,\nonumber\\
G^{+}(x,x^{\prime};E)  &  =\frac{m}{i\hbar^{2}\sqrt{kk_{u}}}\left[
e^{-ikx}t^{\prime}(E)e^{ik_{u}x^{\prime}}+e^{-ikx}r^{\prime}(E)e^{-ik_{u}%
x^{\prime}}\right]  ,x<0,0<x^{\prime}<d,\nonumber\\
G^{+}(x,x^{\prime};E)  &  =\frac{m}{i\hbar^{2}k}\left[  e^{ik\left\vert
x-x^{\prime}\right\vert }+r(E)e^{-ik(x+x^{\prime})}\right]  ,x<0,x^{\prime}<0,
\label{13}%
\end{align}
where the transmission and reflection amplitudes are defined as%
\begin{align}
t(E)  &  =\frac{4\sqrt{kk_{\Delta}}k_{u}e^{ik_{u}d}}{d(E)},t^{\prime}%
(E)=\frac{2\sqrt{kk_{u}}(k_{\Delta}+k_{u})}{d(E)},\nonumber\\
r^{\prime}(E)  &  =\frac{2\sqrt{kk_{u}}(k_{u}-k_{\Delta})e^{2ik_{u}d}}%
{d(E)},r(E)=\frac{(k-k_{u})(k_{\Delta}+k_{u})-(k+k_{u})(k_{\Delta}%
-k_{u})e^{2ik_{u}d}}{d(E)},\nonumber\\
d(E)  &  =(k+k_{u})(k_{\Delta}+k_{u})-(k-k_{u})(k_{\Delta}-k_{u})e^{2ik_{u}d}.
\label{14}%
\end{align}
We remind that $k$, $k_{u}$ and $k_{\Delta}$ are the
perpendicular-to-interface components of the particle wave vector in different
spatial areas which depend on the energy $E$ and $\mathbf{k}_{||}$ as
indicated in (\ref{2}), and $\mathbf{k}_{||}$ is the parallel-to-interface
component of this vector which is conserved for the considered specular
scattering at the interfaces. Using the same approach, it is not difficult to
obtain the Green function $G^{+}(x,x^{\prime};E)$ for other areas of arguments
$x$ and $x^{\prime}$.

The transmission probability $\left\vert t(E)\right\vert ^{2}$ through and
reflection probability $\left\vert r(E)\right\vert ^{2}$ from the asymmetric
potential (\ref{1}), which follow from (\ref{14}) for real $k_{u}$ and
$k_{\Delta}$, are given by
\begin{align}
\left\vert t(E)\right\vert ^{2}  &  =\frac{4kk_{u}^{2}k_{\Delta}}%
{(k+k_{\Delta})^{2}k_{u}^{2}+(k^{2}-k_{u}^{2})(k_{\Delta}^{2}-k_{u}^{2}%
)\sin^{2}(k_{u}d)},\nonumber\\
\left\vert r(E)\right\vert ^{2}  &  =\frac{k_{u}^{2}(k-k_{\Delta})^{2}%
+(k^{2}-k_{u}^{2})(k_{\Delta}^{2}-k_{u}^{2})\sin^{2}(k_{u}d)}{(k+k_{\Delta
})^{2}k_{u}^{2}+(k^{2}-k_{u}^{2})(k_{\Delta}^{2}-k_{u}^{2})\sin^{2}(k_{u}d)}.
\label{14'}%
\end{align}
Note that when $k_{u}d=n\pi$ ($n$ is integer), the resonance transmission
($\left\vert t(E)\right\vert ^{2}=1$ and $\left\vert r(E)\right\vert ^{2}=0$)
happens only for a symmetric rectangular potential with $\Delta=0$
($k_{\Delta}=k$).

In accordance with the obtained results for Green's functions, we will
consider the situation when a particle, given originally by a wave packet
localized to the left of the potential area, i.e. at $x^{\prime}<0$, moves
towards the potential (\ref{1}). We also choose $\Delta\geqslant0$, which
corresponds to the case when, e.g., the spin-up electrons of the left magnetic
layer ($x^{\prime}<0$) move through the nonmagnetic spacer to the right
magnetic layer ($x>d$) aligned either in parallel ($\Delta=0$) or antiparallel
($\Delta>0$) to the left magnetic layer. At the same time, the amplitude $U$
in the potential (\ref{1}) may acquire both positive (barrier) and negative
(well) values.

From Eqs. (\ref{13}) we see that $G^{+}(x,x^{\prime};E)=G^{+}(x^{\prime}%
,x;E)$, and, therefore, the advanced Green function $G^{-}(x,x^{\prime
};E)=\left[  G^{+}(x^{\prime},x;E)\right]  ^{\ast}=\left[  G^{+}(x,x^{\prime
};E)\right]  ^{\ast}$ (see, e.g. \cite{Economou}). Thus, the transmission
amplitude (\ref{4}) is determined by the imaginary part of the Green function
and can be written as%
\begin{equation}
K(x,t;x^{\prime},t_{0})=-\theta(t-t_{0})\frac{1}{\pi}%
{\displaystyle\int\limits_{-\infty}^{\infty}}
dEe^{-\frac{i}{\hbar}E(t-t_{0})}\operatorname{Im}G^{+}(x,x^{\prime};E).
\label{15}%
\end{equation}

Formulas (\ref{13}) - (\ref{15}) present the exact solution for the particle
propagator in the presence of the potential (\ref{1}) in terms of integrals of
elementary functions for a given angle\textbf{\ }($\mathbf{k}_{||}$) of a
particle's arrival at the potential (\ref{1}). Thus the Green function and
propagator are dependent on the additional argument $\mathbf{k}_{||}$, i.e.
actually we have obtained the solution for $G^{+}(x,x^{\prime};E;\mathbf{k}%
_{||})$ and $K(x,t;x^{\prime},t_{0};\mathbf{k}_{||})$. It should be kept in
mind that the wave numbers (\ref{2}) and, therefore, the quantities
$t(E;\mathbf{k}_{||})$, $t^{\prime}(E;\mathbf{k}_{||})$, $r^{\prime
}(E;\mathbf{k}_{||})$, and $r(E;\mathbf{k}_{||})$ in (\ref{14}) are different
in the $%
{\displaystyle\int\limits_{-\infty}^{\hbar^{2}\mathbf{k}_{||}^{2}/2m}}
dE$ and $%
{\displaystyle\int\limits_{\hbar^{2}\mathbf{k}_{||}^{2}/2m}^{\infty}}
dE$ energy integration areas: in the former case, $k(E;\mathbf{k}_{||})$ and
$k_{\Delta}(E;\mathbf{k}_{||})$ ($\Delta\geqslant0$) should be replaced with
$i\overline{k}(E;\mathbf{k}_{||})$ and $i\overline{k}_{\Delta}(E;\mathbf{k}%
_{||})$, where $\overline{k}(E;\mathbf{k}_{||})=\sqrt{\mathbf{k}_{||}%
^{2}-2mE/\hbar^{2}}$ and $\overline{k}_{\Delta}(E;\mathbf{k}_{||}%
)=\sqrt{\mathbf{k}_{||}^{2}+2m(\Delta-E)/\hbar^{2}}$. At the same time, for
energies $E<\hbar^{2}\mathbf{k}_{||}^{2}/2m$, the wave number $k_{u}%
=i\overline{k}_{u}$, $\overline{k}_{u}=\sqrt{\mathbf{k}_{||}^{2}%
+2m(U-E)/h^{2}} $, for $U>0$ (barrier), but for $U<0$ it is real, i.e.
$k_{u}=\sqrt{2m(E+\left\vert U\right\vert )/\hbar^{2}-\mathbf{k}_{||}^{2}}$,
if $E>\hbar^{2}\mathbf{k}_{||}^{2}/2m-\left\vert U\right\vert $ and
$k_{u}=i\overline{k}_{u}$, $\overline{k}_{u}=\sqrt{\mathbf{k}_{||}%
^{2}-2m(E+\left\vert U\right\vert )/\hbar^{2}}$ if $E<\hbar^{2}\mathbf{k}%
_{||}^{2}/2m-\left\vert U\right\vert $. It follows that the "free" Green
function $G_{0}^{+}(x,x^{\prime};E)=\frac{m}{i\hbar^{2}k}e^{ik\left\vert
x-x^{\prime}\right\vert }$ is real in the energy interval ($-\infty\div
\hbar^{2}\mathbf{k}_{||}^{2}/2m$) and, therefore, does not contribute in this
interval to the corresponding "free" propagator $K_{0}(x,t;x^{\prime},t_{0}) $
defined by (\ref{15}). It is also remarkable that for energies $E<\hbar
^{2}\mathbf{k}_{||}^{2}/2m$ the imaginary parts of the Green functions vanish
in all spatial regions, as is seen from definitions (\ref{13}) and (\ref{14})
(e.g., $\operatorname{Im}t(E)=0$ and $\operatorname{Im}r(E)=0$ for
$E<\hbar^{2}\mathbf{k}_{||}^{2}/2m$). Therefore, the energy interval
($-\infty\div\hbar^{2}\mathbf{k}_{||}^{2}/2m$) does not contribute to the
propagation of the particles through the potential well/barrier region.

From Eqs. (\ref{2}), (\ref{13}) and (\ref{14}) we see that the dependence of
the Green function on $E$ and $\mathbf{k}_{||}$ comes in the combination
$E-\hbar^{2}\mathbf{k}_{||}^{2}/2m$, and, therefore, it is convenient to shift
to this new energy variable, which is the perpendicular-to-interface component
of the total particle energy. Thus, accounting for (\ref{13}) - (\ref{15}) and
that for the new energy variable the energy interval ($-\infty\div0$) does not
contribute to the propagator, we have for $t>t_{0}$%
\begin{gather}
K(x,t;x^{\prime},t_{0};\mathbf{k}_{||})=\frac{e^{-\frac{i}{\hbar}\frac
{\hbar^{2}\mathbf{k}_{||}^{2}}{2m}(t-t_{0})}}{\pi\hbar}%
{\displaystyle\int\limits_{0}^{\infty}}
dEe^{-\frac{i}{\hbar}E(t-t_{0})}\frac{1}{\sqrt{v(E)}}\operatorname{Re}\left[
\frac{1}{\sqrt{v_{\Delta}(E)}}t(E)e^{ik_{\Delta}(E)(x-d)}e^{-ik(E)x^{\prime}%
}\right]  ,\nonumber\\
x^{\prime}<0,x>d,\nonumber\\
K(x,t;x^{\prime},t_{0};\mathbf{k}_{||})=\frac{e^{-\frac{i}{\hbar}\frac
{\hbar^{2}\mathbf{k}_{||}^{2}}{2m}(t-t_{0})}}{\pi\hbar}%
{\displaystyle\int\limits_{0}^{\infty}}
dEe^{-\frac{i}{\hbar}E(t-t_{0})}\frac{1}{\sqrt{v(E)}}\operatorname{Re}\left\{
\frac{e^{-ik(E)x^{\prime}}}{\sqrt{v_{u}(E)}}\left[  t^{\prime}(E)e^{ik_{u}%
(E)x}+r^{\prime}(E)e^{-ik_{u}(E)x}\right]  \right\}  ,\nonumber\\
x^{\prime}<0,0<x<d,\nonumber\\
K(x,t;x^{\prime},t_{0};\mathbf{k}_{||})=\frac{e^{-\frac{i}{\hbar}\frac
{\hbar^{2}\mathbf{k}_{||}^{2}}{2m}(t-t_{0})}}{\pi\hbar}%
{\displaystyle\int\limits_{0}^{\infty}}
dEe^{-\frac{i}{\hbar}E(t-t_{0})}\frac{1}{v(E)}\operatorname{Re}%
[e^{ik(E)\left\vert x-x^{\prime}\right\vert }+r(E)e^{-ik(E)(x+x^{\prime}%
)}],x^{\prime}<0,x<0, \label{16}%
\end{gather}
where (see (\ref{2}))
\begin{align}
v(E)  &  =\hbar k(E)/m,k(E)=\sqrt{2mE/\hbar^{2}},\nonumber\\
v_{\Delta}(E)  &  =\hbar k_{\Delta}(E)/m,k_{\Delta}(E)=\sqrt{2m(E-\Delta
)/\hbar^{2}},\nonumber\\
v_{u}(E)  &  =\hbar k_{u}(E)/m,k_{u}(E)=\sqrt{2m(E-U)/\hbar^{2}}. \label{17}%
\end{align}
The transmission and reflection amplitudes $t(E)$, $r(E)$, $t^{\prime}(E)$ and
$r^{\prime}(E)$ in (\ref{16}) are defined by (\ref{14}) with the wave numbers
(\ref{17}).

It is easy to verify that the integration over $E$ and $\mathbf{k}_{||}$
(according to (\ref{6})) of the first term in the last line of (\ref{16})
results in the known formula for the space-time propagator for a freely moving
particle
\begin{equation}
K_{0}(\mathbf{r},t;\mathbf{r}^{\prime},t_{0})=\theta(t-t_{0})\left[  \frac
{m}{2\pi i\hbar(t-t_{0})}\right]  ^{3/2}\exp\left[  \frac{im(\mathbf{r}%
-\mathbf{r}^{\prime})^{2}}{2\hbar(t-t_{0})}\right]  ,x<0,x^{\prime}<0.
\label{17a}%
\end{equation}

The obtained results for the particle propagator completely resolve (by means
of Eq. (\ref{3})) the time-dependent Schr\"{o}dinger equation for a particle
moving under the influence of the rectangular potential (\ref{1}).

\section{Time-dependent probability density of finding a particle in different
spatial regions}

Using Eqs. (\ref{3}), (\ref{6}), (\ref{13}) and (\ref{15}), we can present the
wave function in different spatial regions at $t>t_{0}$ as%
\begin{equation}
\psi(\mathbf{r,}t)=\psi_{>}(\mathbf{r,}t)+\psi_{<}(\mathbf{r,}t). \label{18}%
\end{equation}
Here%
\begin{align}
\psi_{>}(\mathbf{r},t)  &  =\frac{1}{\sqrt{2\pi\hbar}}%
{\displaystyle\int\limits_{0}^{\infty}}
dEe^{-\frac{i}{\hbar}E(t-t_{0})}\frac{1}{\sqrt{v_{\Delta}(E)}}%
t(E)e^{ik_{\Delta}(E)(x-d)}\psi_{>}(E;\mathbf{\rho,}t),x>d,\nonumber\\
\psi_{<}(\mathbf{r},t)  &  =\frac{1}{\sqrt{2\pi\hbar}}%
{\displaystyle\int\limits_{0}^{\infty}}
dEe^{-\frac{i}{\hbar}E(t-t_{0})}\frac{1}{\sqrt{v_{\Delta}^{\ast}(E)}}t^{\ast
}(E)e^{-ik_{\Delta}^{\ast}(E)(x-d)}\psi_{<}(E;\mathbf{\rho,}t),x>d,\nonumber\\
\psi_{>}(\mathbf{r},t)  &  =\frac{1}{\sqrt{2\pi\hbar}}%
{\displaystyle\int\limits_{0}^{\infty}}
dEe^{-\frac{i}{\hbar}E(t-t_{0})}\frac{1}{\sqrt{v_{u}(E)}}[t^{\prime
}(E)e^{ik_{u}(E)x}+r^{\prime}(E)e^{-ik_{u}(E)x}]\psi_{>}(E;\mathbf{\rho
,}t),0<x<d,\nonumber\\
\psi_{<}(\mathbf{r},t)  &  =\frac{1}{\sqrt{2\pi\hbar}}%
{\displaystyle\int\limits_{0}^{\infty}}
dEe^{-\frac{i}{\hbar}E(t-t_{0})}\frac{1}{\sqrt{v_{u}^{\ast}(E)}}[t^{\prime
\ast}(E)e^{-ik_{u}^{\ast}(E)x}+r^{\prime\ast}(E)e^{ik_{u}^{\ast}(E)x}]\psi
_{<}(E;\mathbf{\rho,}t),0<x<d,\nonumber\\
\psi_{>}(\mathbf{r},t)  &  =\frac{1}{\sqrt{2\pi\hbar}}%
{\displaystyle\int\limits_{0}^{\infty}}
dEe^{-\frac{i}{\hbar}E(t-t_{0})}\frac{1}{\sqrt{v(E)}}[e^{ik(E)x}%
+r(E)e^{-ik(E)x}]\psi_{>}(E;\mathbf{\rho,}t),x<0,\nonumber\\
\psi_{<}(\mathbf{r},t)  &  =\frac{1}{\sqrt{2\pi\hbar}}%
{\displaystyle\int\limits_{0}^{\infty}}
dEe^{-\frac{i}{\hbar}E(t-t_{0})}\frac{1}{\sqrt{v(E)}}[e^{-ik(E)x}+r^{\ast
}(E)e^{ik(E)x}]\psi_{<}(E;\mathbf{\rho,}t),x<0 \label{19}%
\end{align}
and $\mathbf{r=(}x,\mathbf{\rho)}$. The wave function in the $E$%
-representation $\psi_{>(<)}(E;\mathbf{\rho,}t)$ is related to its
$k$-representation $\psi_{>(<)}[k(E);\mathbf{\rho,}t)]$ as
\begin{align}
\psi_{>}(E;\mathbf{\rho,}t)  &  =\frac{1}{\sqrt{2\pi\hbar v(E)}}\psi
_{>}[k(E);\mathbf{\rho,}t],\psi_{<}(E;\mathbf{\rho,}t)=\frac{1}{\sqrt
{2\pi\hbar v(E)}}\psi_{<}[k(E);\mathbf{\rho,}t],\nonumber\\
\psi_{>}[k(E);\mathbf{\rho,}t]  &  =\int d\mathbf{\rho}^{\prime}%
K(\mathbf{\rho,}t;\mathbf{\rho}^{\prime}\mathbf{,}t_{0})%
{\displaystyle\int}
dx^{\prime}e^{-ik(E)x^{\prime}}\psi(x^{\prime},\mathbf{\rho}^{\prime}%
,t_{0}),\nonumber\\
\psi_{<}[k(E);\mathbf{\rho},t]  &  =\int d\mathbf{\rho}^{\prime}%
K(\mathbf{\rho,}t;\mathbf{\rho}^{\prime}\mathbf{,}t_{0})%
{\displaystyle\int}
dx^{\prime}e^{ik(E)x^{\prime}}\psi(x^{\prime},\mathbf{\rho}^{\prime}%
,t_{0}),\nonumber\\
K(\mathbf{\rho,}t;\mathbf{\rho}^{\prime}\mathbf{,}t_{0})  &  =\frac{1}{A}%
\sum\limits_{\mathbf{k}_{||}}\exp[\mathbf{-}\frac{i}{\hbar}\frac{\hbar
^{2}\mathbf{k}_{||}^{2}}{2m}(t-t_{0})]e^{i\mathbf{k}_{||}(\mathbf{\rho-\rho
}^{\prime})}=\frac{m}{2\pi i\hbar(t-t_{0})}\exp[-\frac{(\mathbf{\rho-\rho
}^{\prime})^{2}m}{2i\hbar(t-t_{0})}], \label{20}%
\end{align}
where $K(\mathbf{\rho,}t;\mathbf{\rho}^{\prime}\mathbf{,}t_{0})$ is the "free"
propagator in the parallel-to-interface ($y,z$) plane (see (\ref{6}),
(\ref{16}) and (\ref{17a})).

It can be verified that the wave functions (\ref{19}) and their derivatives
are continuous at $x=0$ and $x=d$. To be definite, we assume that for positive
energies $k(E)=\sqrt{2mE/\hbar^{2}}>0$ and, therefore, $\psi_{>}%
[k(E);\mathbf{\rho},t]$ is related to the component of the initial wave
function $\psi(x^{\prime},\mathbf{\rho}^{\prime},t_{0})$ corresponding to
propagation to the right along the $x$ axis, and, accordingly, $\psi
_{<}[k(E);\mathbf{\rho},t]$ represents propagation to the left. When the
potential $V(x)\neq0$, integration over $x^{\prime}$ in (\ref{20}) is
restricted to the negative semispace ($x^{\prime}<0$), as it follows from the
expressions (\ref{16}) for the particle propagator.

The result, given by Eqs. (\ref{18}), (\ref{19}) and (\ref{20}), indicates
that, generally, the contribution of the wave function, originated at
$t=t_{0}$ to the left of the potential (\ref{1}) ($x^{\prime}<0$), to the wave
function in the region of the potential ($0<x<d$) and to the right of it
($x>d$) comes at $t>t_{0}$ from both: the components moving to the right,
$\psi_{>}$, and to the left, $\psi_{<}$. This rather paradoxical result
follows from the fact that if the initial wave packet has the non-negligible
negative momentum components (restricted to a half line in the momentum
space), the corresponding spatial wave function is different from zero in the
entire $x$-region ($-\infty,\infty$), interacting with the potential even at
$t<t_{0}$, and is thus modified by this interaction (see also \cite{Muga et al
1992}, \cite{Muga Sources 2001}). As a result, the backward-moving components
contribute to the behavior of the wave function at $t>t_{0}$ in the spatial
regions to the right of the original wave packet localization.

Consequently, the probability density of finding a particle in the spacetime
point ($\mathbf{r},t$), $\left\vert \psi(\mathbf{r},t)\right\vert ^{2}$ is
determined by the forward- and backward-moving terms, as well as their
interference:%
\begin{equation}
\left\vert \psi(\mathbf{r},t)\right\vert ^{2}=\left\vert \psi_{>}%
(\mathbf{r},t)\right\vert ^{2}+\left\vert \psi_{<}(\mathbf{r},t)\right\vert
^{2}+2\operatorname{Re}\psi_{>}(\mathbf{r},t)\psi_{<}^{\ast}(\mathbf{r},t).
\label{21}%
\end{equation}
Equations (\ref{19}) - (\ref{21}) generally resolve the problem of finding a
particle in the spatial region of interest at time $t$ for a given initial
wave function $\psi(\mathbf{r}^{\prime},t_{0})$. These equations can be used
for numerical modeling of the corresponding probability density in the
different space-time regions (see below) and for determining some
characteristics of the particle dynamics under the influence of the potential
(\ref{1}).

In order to estimate the actual contribution of the backward-moving and
interference terms to the obtained general formulas, we should consider a
physically relevant situation as to the initial wave packet. Let us consider
the case when the moving particles are associated with a wave packet which is
initially sufficiently well localized to the left of the potential (\ref{1}).
Thus we now consider the problem for a particular case of the initial state
corresponding to the wave packet%
\begin{equation}
\psi(\mathbf{r}^{\prime},t_{0})=\frac{1}{(2\pi\sigma^{2})^{3/4}}\exp\left[
-\frac{(\mathbf{r}^{\prime}-\mathbf{r}_{i})^{2}}{4\sigma^{2}}+i\mathbf{k}%
_{i}\mathbf{r}^{\prime}\right]  ,x_{i}<0,k_{i}>0, \label{22}%
\end{equation}
located in the vicinity of $\mathbf{r}_{i}=(x_{i},\mathbf{\rho}_{i})$ and
moving in the positive $x$ direction with the average momentum $\mathbf{p}%
_{i}=\hbar\mathbf{k}_{i}$, $k_{i}=k_{x}>0$ ($\mathbf{k}_{i}=(k_{i}%
,\mathbf{k}_{||}^{i}),\mathbf{r}^{\prime}=(x^{\prime},\mathbf{\rho}^{\prime}%
)$). Thus, we consider a general situation, when a particle, associated with
the wave packet (\ref{22}), comes to the potential (\ref{1}) from the left
with the positive perpendicular-to-interface momentum component $\hbar
k_{i}>0$ at the angle defined by the parallel-to-inteface momentum component
$\hbar\mathbf{k}_{||}^{i}$. Now, we can perform integration over spatial
variables $x^{\prime}$, $\mathbf{\rho}^{\prime}$, as it follows from
(\ref{20}) and (\ref{22}).The result is
\begin{align}
\psi_{>}(E;\mathbf{\rho,}t)  &  =C_{||}(\mathbf{\rho,}t)\psi_{>}(E),\psi
_{<}(E;\mathbf{\rho,}t)=C_{||}(\mathbf{\rho,}t)\psi_{<}(E),\nonumber\\
\psi_{>}(E)  &  =\frac{1}{\sqrt{\pi\hbar v(E)}}(2\pi\sigma^{2})^{1/4}%
e^{i[k_{i}-k(E)]x_{i}}e^{-[k_{i}-k(E)]^{2}\sigma^{2}},\nonumber\\
\psi_{<}(E)  &  =\frac{1}{\sqrt{\pi\hbar v(E)}}(2\pi\sigma^{2})^{1/4}%
e^{i[k_{i}+k(E)]x_{i}}e^{-[k_{i}+k(E)]^{2}\sigma^{2}},\nonumber\\
C_{||}(\mathbf{\rho,}t)  &  =\sqrt{\frac{2}{\pi}}\frac{m\sigma}{i\hbar
(t-t_{0})+2m\sigma^{2}}\exp[-\frac{(\mathbf{\rho}-\mathbf{\rho}_{i}%
-2i\mathbf{k}_{||}^{i}\sigma^{2})^{2}m}{2i\hbar(t-t_{0})+4m\sigma^{2}%
}]e^{i\mathbf{k}_{||}^{i}\mathbf{\rho}_{i}}e^{-(\mathbf{k}_{||}^{i}\sigma
)^{2}}, \label{23}%
\end{align}
\qquad where the factor $C_{||}(\mathbf{\rho,}t)$ defines the dependence on
the parallel-to-interface components of the vectors involved. Thus, the
forward- and backward-moving components of the wave function $\psi
_{>(<)}(\mathbf{r,}t)$ (\ref{19}) for the initial wave packet (\ref{22})
reduce to the one-dimensional integral over energy $E$ with the
energy-dependent functions $\psi_{>(<)}(E)$ and the common factor
$C_{||}(\mathbf{\rho,}t)$.

We note that
\begin{equation}
\int d\mathbf{\rho}\left\vert C_{||}(\mathbf{\rho,}t)\right\vert ^{2}=1,
\label{23a}%
\end{equation}
and, therefore, the total probability density of finding a particle in the
given space-time point ($x,t$)
\begin{align}
\left\vert \psi(x,t)\right\vert ^{2}  &  =\int d\mathbf{\rho}\left\vert
\psi(\mathbf{r},t)\right\vert ^{2}\nonumber\\
&  =\left\vert \psi_{>}(x,t)\right\vert ^{2}+\left\vert \psi_{<}%
(x,t)\right\vert ^{2}+2\operatorname{Re}\psi_{>}(x,t)\psi_{<}^{\ast}(x,t),
\label{23b}%
\end{align}
as it follows from (\ref{21}), and the functions $\psi_{>(<)}(x,t)$ are
determined by Eqs. (\ref{19}) where $\psi_{>(<)}(E;\mathbf{\rho,}t)$ is
replaced with $\psi_{>(<)}(E)$ (see (\ref{23})).

A physically relevant situation occurs when the initial wave function vanishes
at $x>0$ (well localized within the $x<0$ half-line) because the propagator
(\ref{16}) transmits this function from the $x^{\prime}<0$ region to the $x>0$
or $x<0$ regions. This can be achieved if we define the initial wave function
as (\ref{22}) at $x^{\prime}<0$ and set it zero at $x^{\prime}>0$. It can be
shown that when the condition
\begin{equation}
\left\vert \frac{x_{i}}{2\sigma}\right\vert \gg1 \label{24}%
\end{equation}
holds (i.e. when the tail of the initial wave packet (\ref{22}) is very small
near the arrival point $x=0$), the Fourier transform of the initial wave
packet matches the Fourier transform of a cutoff Gaussian wave packet, defined
as (\ref{22}) at $x^{\prime}<0$ and zero at $x^{\prime}>0$ (see \cite{Cordero
2010}).

Generally, both the $\psi_{>}(\mathbf{r},t)$ and $\psi_{<}(\mathbf{r},t)$
components contribute to the probability density $\left\vert \psi
(\mathbf{r},t)\right\vert ^{2}$ (see (\ref{21})). We can also assume that
\begin{equation}
k_{i}\sigma\gg1, \label{25}%
\end{equation}
which implies that the perpendicular-to-interface momentum dispersion
$\hbar/2\sigma$ is much smaller than the corresponding characteristic momentum
$p_{i}=\hbar k_{i}$, or, equivalently,
\begin{align}
\frac{\hbar^{2}}{2m\sigma^{2}}  &  \ll E_{\bot},\nonumber\\
E_{\bot}  &  =\frac{\hbar^{2}k_{i}^{2}}{2m}=E_{i}-\frac{\hbar^{2}%
\mathbf{k}_{||}^{i2}}{2m}, \label{26}%
\end{align}
i.e., the energy dispersion $\hbar^{2}/8m\sigma^{2}$ is much smaller than the
perpendicular component $E_{\bot}$ of the incident particle energy
$E_{i}=(\hbar^{2}/2m)(k_{i}^{2}+\mathbf{k}_{||}^{i2})$. Then one can see from
(\ref{19}) and (\ref{23}) that in the case when condition (\ref{25}) holds,
the contribution of the backward-moving term $\psi_{<}(\mathbf{r},t)$ to the
probability density is significantly smaller than that of the forward-moving
term $\psi_{>}(\mathbf{r},t)$, and, therefore, in the first approximation the
former can be neglected. Thus, the backward-moving term $\psi_{<}%
(\mathbf{r},t)$ is not essential in the quasi-classical approximation when
both inequalities (\ref{24}) and (\ref{25}) are satisfied and, therefore, the
particle scattering at the potential (\ref{1}) is associated with the wave
packet (\ref{22}) characterized by a well-defined location relative to the
potential and well-defined momentum. However, if the inequality (\ref{25}) (or
(\ref{26})) is violated, then both the forward- and backward-moving components
of the wave function (\ref{19}) equally contribute to the probability density
$\left\vert \psi(\mathbf{r,}t)\right\vert ^{2}$. In this case the
quasi-classical approximation is not relevant and the particle is associated
with the well-localized wave packet which has the broad
perpendicular-to-interface momentum (energy) distribution.

\section{Stationary case and numerical modeling}

We will consider the probability density $\left\vert \psi(x\mathbf{,}%
t)\right\vert ^{2}$ (\ref{23b}). It is convenient to shift to dimensionless
variables. As seen from (\ref{19}), there is a natural spatial scale $d$, an
energy scale $E_{d}=\hbar^{2}/2md^{2}$ (the energy uncertainty due to particle
localization within a barrier of width $d$), and a corresponding time scale
$t_{d}=\hbar/E_{d}$. Then, using (\ref{19}) and (\ref{23}) (with
$C_{||}(\mathbf{\rho},t)=1$), we can obtain the wave function $\psi(x,t)$ in
the different spatial regions resulting from the evolution of the initial
Gaussian wave packet (\ref{22}) in the presence of the potential barrier
(\ref{1}). Thus, the one-dimensional wave function, following from (\ref{19})
and needed for the calculation of the probability density (\ref{23b}), in the
dimensionless variables is%
\begin{gather}
\psi(\widetilde{x},\widetilde{t})=\psi_{>}(\widetilde{x},\widetilde{t}%
)+\psi_{<}(\widetilde{x},\widetilde{t}),\widetilde{t}>\widetilde{t}%
_{0},\nonumber\\
\psi_{>}(\widetilde{x},\widetilde{t})=\frac{1}{\sqrt{2}(2\pi)^{3/4}}%
\frac{\widetilde{\sigma}^{1/2}}{d^{1/2}}e^{ik_{i}x_{i}}%
{\displaystyle\int\limits_{0}^{\infty}}
\frac{d\widetilde{E}}{\widetilde{E}^{1/4}(\widetilde{E}-\widetilde{\Delta
})^{1/4}}e^{-i\widetilde{E}(\widetilde{t}-\widetilde{t}_{0})}t(\widetilde
{E})\nonumber\\
\times e^{-(\sqrt{\widetilde{E}}-\sqrt{\widetilde{E}}_{\perp})^{2}%
\widetilde{\sigma}^{2}}e^{i\sqrt{\widetilde{E}-\widetilde{\Delta}}%
(\widetilde{x}-1)}e^{-i\sqrt{\widetilde{E}}\widetilde{x}_{i}},\widetilde
{x}>1,\nonumber\\
\psi_{<}(\widetilde{x},\widetilde{t})=\frac{1}{\sqrt{2}(2\pi)^{3/4}}%
\frac{\widetilde{\sigma}^{1/2}}{d^{1/2}}e^{ik_{i}x_{i}}%
{\displaystyle\int\limits_{0}^{\infty}}
\frac{d\widetilde{E}}{\widetilde{E}^{1/4}[(\widetilde{E}-\widetilde{\Delta
})^{1/4}]^{\ast}}e^{-i\widetilde{E}(\widetilde{t}-\widetilde{t}_{0})}t^{\ast
}(\widetilde{E})\nonumber\\
e^{-(\sqrt{\widetilde{E}}+\sqrt{\widetilde{E}}_{\perp})^{2}\widetilde{\sigma
}^{2}}e^{-i(\sqrt{\widetilde{E}-\widetilde{\Delta}})^{\ast}(\widetilde{x}%
-1)}e^{i\sqrt{\widetilde{E}}\widetilde{x}_{i}},\widetilde{x}>1,\nonumber\\
\psi_{>}(\widetilde{x},\widetilde{t})=\frac{1}{\sqrt{2}(2\pi)^{3/4}}%
\frac{\widetilde{\sigma}^{1/2}}{d^{1/2}}e^{ik_{i}x_{i}}%
{\displaystyle\int\limits_{0}^{\infty}}
\frac{d\widetilde{E}}{\widetilde{E}^{1/4}(\widetilde{E}-\widetilde{U})^{1/4}%
}e^{-i\widetilde{E}(\widetilde{t}-\widetilde{t}_{0})}[t^{\prime}(\widetilde
{E})e^{i\sqrt{\widetilde{E}-\widetilde{U}}\widetilde{x}}\nonumber\\
+r^{\prime}(\widetilde{E})e^{-i\sqrt{\widetilde{E}-\widetilde{U}}\widetilde
{x}}]e^{-(\sqrt{\widetilde{E}}-\sqrt{\widetilde{E}}_{\perp})^{2}%
\widetilde{\sigma}^{2}}e^{-i\sqrt{\widetilde{E}}\widetilde{x}_{i}%
},0<\widetilde{x}<1,\nonumber\\
\psi_{<}(\widetilde{x},\widetilde{t})=\frac{1}{\sqrt{2}(2\pi)^{3/4}}%
\frac{\widetilde{\sigma}^{1/2}}{d^{1/2}}e^{ik_{i}x_{i}}%
{\displaystyle\int\limits_{0}^{\infty}}
\frac{d\widetilde{E}}{\widetilde{E}^{1/4}[(\widetilde{E}-\widetilde{U}%
)^{1/4}]^{\ast}}e^{-i\widetilde{E}(\widetilde{t}-\widetilde{t}_{0})}%
[t^{\prime}(\widetilde{E})e^{i\sqrt{\widetilde{E}-\widetilde{U}}\widetilde{x}%
}\nonumber\\
+r^{\prime}(\widetilde{E})e^{-i\sqrt{\widetilde{E}-\widetilde{U}}\widetilde
{x}}]^{\ast}e^{-(\sqrt{\widetilde{E}}+\sqrt{\widetilde{E}}_{\perp}%
)^{2}\widetilde{\sigma}^{2}}e^{i\sqrt{\widetilde{E}}\widetilde{x}_{i}%
},0<\widetilde{x}<1,\nonumber\\
\psi_{>}(\widetilde{x},\widetilde{t})=\frac{1}{\sqrt{2}(2\pi)^{3/4}}%
\frac{\widetilde{\sigma}^{1/2}}{d^{1/2}}e^{ik_{i}x_{i}}%
{\displaystyle\int\limits_{0}^{\infty}}
\frac{d\widetilde{E}}{\sqrt{\widetilde{E}}}e^{-i\widetilde{E}(\widetilde
{t}-\widetilde{t}_{0})}[e^{i\sqrt{\widetilde{E}}\widetilde{x}}\nonumber\\
+r(\widetilde{E})e^{-i\sqrt{\widetilde{E}}\widetilde{x}}]e^{-(\sqrt
{\widetilde{E}}-\sqrt{\widetilde{E}}_{\perp})^{2}\widetilde{\sigma}^{2}%
}e^{-i\sqrt{\widetilde{E}}\widetilde{x}_{i}},\widetilde{x}<0,\nonumber\\
\psi_{<}(\widetilde{x},\widetilde{t})=\frac{1}{\sqrt{2}(2\pi)^{3/4}}%
\frac{\widetilde{\sigma}^{1/2}}{d^{1/2}}e^{ik_{i}x_{i}}%
{\displaystyle\int\limits_{0}^{\infty}}
\frac{d\widetilde{E}}{\sqrt{\widetilde{E}}}e^{-i\widetilde{E}(\widetilde
{t}-\widetilde{t}_{0})}[e^{-i\sqrt{\widetilde{E}}\widetilde{x}}\nonumber\\
+r^{\ast}(\widetilde{E})e^{i\sqrt{\widetilde{E}}\widetilde{x}}]e^{-(\sqrt
{\widetilde{E}}+\sqrt{\widetilde{E}}_{\perp})^{2}\widetilde{\sigma}^{2}%
}e^{i\sqrt{\widetilde{E}}\widetilde{x}_{i}},\widetilde{x}<0, \label{27}%
\end{gather}
where
\begin{align}
t(\widetilde{E})  &  =\frac{4\widetilde{E}^{1/4}\sqrt{\widetilde{E}%
-\widetilde{U}}(\widetilde{E}-\widetilde{\Delta})^{1/4}e^{i\sqrt{\widetilde
{E}-\widetilde{U}}}}{d(\widetilde{E})},\nonumber\\
t^{\prime}(\widetilde{E})  &  =\frac{2\widetilde{E}^{1/4}(\widetilde
{E}-\widetilde{U})^{1/4}(\sqrt{\widetilde{E}-\widetilde{\Delta}}%
+\sqrt{\widetilde{E}-\widetilde{U}})}{d(\widetilde{E})},\nonumber\\
r^{\prime}(\widetilde{E})  &  =\frac{2\widetilde{E}^{1/4}(\widetilde
{E}-\widetilde{U})^{1/4}(\sqrt{\widetilde{E}-\widetilde{U}}-\sqrt
{\widetilde{E}-\widetilde{\Delta}})e^{2i\sqrt{\widetilde{E}-\widetilde{U}}}%
}{d(\widetilde{E})},\nonumber\\
r(\widetilde{E})  &  =\frac{%
\begin{array}
[c]{c}%
(\sqrt{\widetilde{E}}-\sqrt{\widetilde{E}-\widetilde{U}})(\sqrt{\widetilde
{E}-\widetilde{\Delta}}+\sqrt{\widetilde{E}-\widetilde{U}})\\
-(\sqrt{\widetilde{E}}+\sqrt{\widetilde{E}-\widetilde{U}})(\sqrt{\widetilde
{E}-\widetilde{\Delta}}-\sqrt{\widetilde{E}-\widetilde{U}})e^{2i\sqrt
{\widetilde{E}-\widetilde{U}}}%
\end{array}
}{d(\widetilde{E})},\nonumber\\
d(\widetilde{E})  &  =(\sqrt{\widetilde{E}}+\sqrt{\widetilde{E}-\widetilde{U}%
})(\sqrt{\widetilde{E}-\widetilde{\Delta}}+\sqrt{\widetilde{E}-\widetilde{U}%
})\nonumber\\
&  -(\sqrt{\widetilde{E}}-\sqrt{\widetilde{E}-\widetilde{U}})(\sqrt
{\widetilde{E}-\widetilde{\Delta}}-\sqrt{\widetilde{E}-\widetilde{U}%
})e^{2i\sqrt{\widetilde{E}-\widetilde{U}}}, \label{28}%
\end{align}
and $\widetilde{E}=E/E_{d}$, $\widetilde{U}=U/E_{d}$, $\widetilde{\Delta
}=\Delta/E_{d}$, $\widetilde{E}_{\bot}=E_{\bot}/E_{d}$, $E_{\bot}=\hbar
^{2}k_{i}^{2}/2m$, $\widetilde{t}=t/t_{d}$, $\widetilde{t}_{0}=t_{0}/t_{d}$,
$\widetilde{\sigma}=\sigma/d$, $\widetilde{x}=x/d$, $\widetilde{x}_{i}%
=x_{i}/d$. The conditions (\ref{24}) and (\ref{26}) read in the dimensionless
variables, correspondingly,%
\begin{equation}
\left\vert \widetilde{x}_{i}\right\vert \gg2\widetilde{\sigma},\widetilde
{E}_{\bot}\gg1/\widetilde{\sigma}^{2}. \label{29}%
\end{equation}

It is instructive to consider first the limiting case defined by the second
inequality (\ref{29}). In this case, the forward-moving terms $\psi
_{>}(\widetilde{x},\widetilde{t})$ in Eqs. (\ref{27}) give the main
contribution to the total wave function, i.e., $\psi(\widetilde{x}%
,\widetilde{t})\thickapprox\psi_{>}(\widetilde{x},\widetilde{t})$. Also, the
integrals over energy in $\psi_{>}(\widetilde{x},\widetilde{t})$ (\ref{27})
can be asymptotically evaluated at $\lambda=\widetilde{E}_{i}\widetilde
{\sigma}^{2}\gg1$\ due to the fact that the contribution to these integrals
mainly comes from the energy region $\widetilde{E}\thickapprox\widetilde
{E}_{\perp}$. In this case, the wave functions $\psi_{>}(\widetilde
{x},\widetilde{t})$ reduce (in the first approximation with $\frac
{1}{\widetilde{E}_{\bot}\widetilde{\sigma}^{2}}\ll1$) to the stationary (for
$\widetilde{E}=\widetilde{E}_{\bot}$) results, oscillating with time as
$\exp[-i\widetilde{E}_{\bot}(\widetilde{t}-\widetilde{t}_{0})]$. Thus, if we
present Eqs. (\ref{27}) for $\psi_{>}(\widetilde{x},\widetilde{t})$ as%
\begin{align}
\psi_{>}(\widetilde{x},\widetilde{t})  &  =\int\limits_{0}^{\infty}%
\varphi(\widetilde{x},\widetilde{x}_{i};\widetilde{E})\exp[-i\widetilde
{E}(\widetilde{t}-\widetilde{t}_{0})]\exp[\lambda f(\widetilde{E}%
)]d\widetilde{E},\nonumber\\
\lambda &  =\widetilde{E}_{\bot}\widetilde{\sigma}^{2}\gg1,f(\widetilde
{E})=-(\sqrt{\widetilde{E}}-\sqrt{\widetilde{E}_{\bot}})^{2}/\widetilde
{E}_{\bot}, \label{30}%
\end{align}
where $\varphi(\widetilde{x},\widetilde{x}_{i};\widetilde{E})$ stands for any
integrand in (\ref{27}) multiplied by exponentials of (\ref{30}), the
asymptotic value of (\ref{30}) is%
\begin{equation}
\psi_{>}(\widetilde{x},\widetilde{t})\backsim\frac{2\sqrt{\pi}}{\widetilde
{\sigma}}\sqrt{\widetilde{E}_{\bot}}\varphi(\widetilde{x},\widetilde{x}%
_{i};\widetilde{E}_{\bot})\exp[-i\widetilde{E}_{\bot}(\widetilde{t}%
-\widetilde{t}_{0})]. \label{31}%
\end{equation}
Accordingly, this stationary result leads to the square modulus of the wave
function $\left\vert \psi_{>}(\widetilde{x},\widetilde{t})\right\vert ^{2}$,
defined by Eqs. (\ref{27}), which is independent of time. For the case of the
potential well ($U<0$) as well as for the potential barrier ($U>0$), we obtain
at $E_{\bot}>\Delta$ (in the original non-scaled variables)
\begin{align}
\left\vert \psi_{>}(x)\right\vert ^{2}  &  =\frac{1}{\sqrt{2\pi}\sigma}%
\frac{16E_{\bot}\left\vert E_{\bot}-U\right\vert }{\left\vert d(E_{\bot
})\right\vert ^{2}},x>d,\nonumber\\
\left\vert \psi_{>}(x)\right\vert ^{2}  &  =\frac{1}{\sqrt{2\pi}\sigma}%
\frac{16E_{\bot}}{\left\vert d(E_{\bot})\right\vert ^{2}}\left\vert E_{\bot
}-\Delta-(U-\Delta)\cos^{2}[\sqrt{2m(E_{\bot}-U)/\hbar^{2}}(x-d)]\right\vert
,0<x<d,\nonumber\\
\left\vert d(E_{\bot})\right\vert ^{2}  &  =\left\vert 4(\sqrt{E_{\bot}}%
+\sqrt{E_{\bot}-\Delta})^{2}(E_{\bot}-U)+4U(U-\Delta)\sin^{2}[\sqrt
{2m(E_{\bot}-U)/\hbar^{2}}d]\right\vert ,\nonumber\\
E_{\bot}  &  \gg\hbar^{2}/2m\sigma^{2}. \label{32}%
\end{align}
Note that when a particle tunnels through a barrier ($E_{\bot}>0$, $U>0$,
$E_{\bot}<U$), $\cos[\sqrt{2m(E_{\bot}-U)/\hbar^{2}}(x-d)]$ and $\sin
[\sqrt{2m(E_{\bot}-U)/\hbar^{2}}d]$ in (\ref{32}) should be replaced with
$\cosh[\sqrt{2m(U-E_{\bot})/\hbar^{2}}(x-d)]$ and $i\sinh[\sqrt{2m(U-E_{\bot
})/\hbar^{2}}d]$, respectively.

Formulae (\ref{32}) provide the spatial dependence of the wave function square
modulus at different spatial regions relative to the potential area for the
stationary case, when the initial wave packet (\ref{22}) is characterized by
an extra narrow distribution in the energy (perpendicular-to-interface
momentum) space. Thus, in this approximation, the transmitted probability
density ($x>d$) is constant in space, while in the potential region ($0<x<d$)
we have the oscillating interference pattern (for $E_{\bot}>U$).

The picture before the potential ($x<0$) is more complicated and results from
the interference of the incoming and reflected waves. The corresponding
formula becomes simplified for the resonant case, when $\sqrt{2m(E_{\bot
}-U)/\hbar^{2}}d=\pi n$ ($n$ is the integer), and is given by ($E_{\bot}>U, $
$E_{\bot}>\Delta$)%
\begin{align}
\left\vert \psi_{>}(x)\right\vert ^{2}  &  =\frac{1}{\sqrt{2\pi}\sigma}%
\frac{4}{(\sqrt{E_{\bot}}+\sqrt{E_{\bot}-\Delta})^{2}}[E_{\bot}-\Delta\sin
^{2}(\sqrt{2mE_{\bot}/\hbar^{2}}x)],x<0,\nonumber\\
E_{\bot}  &  \gg\hbar^{2}/2m\sigma^{2},\sqrt{2m(E_{\bot}-U)/\hbar^{2}}d=\pi n.
\label{33}%
\end{align}
The oscillating interference picture given by (\ref{33}) is caused by the
earlier-mentioned fact that in the case of an asymmetric potential
($\Delta\neq0$), the reflection amplitude $r(E)\neq0$ for the resonant
energies $E$ (see (\ref{14})). From Eqs. (\ref{32}) and (\ref{33}) we see that
the norm $\left\vert \psi_{>}(0)\right\vert ^{2}=\frac{1}{\sqrt{2\pi}\sigma
}\frac{4E_{\bot}}{(\sqrt{E_{\bot}}+\sqrt{E_{\bot}-\Delta})^{2}}$ at the
potential left boundary $x=0$ is transmitted at the resonance condition
$\sqrt{2m(E_{\bot}-U)/\hbar^{2}}d=\pi n$ to the region $x>d$ beyond the
potential. Only for a symmetric rectangular potential ($\Delta=0$) the
reflection amplitude $r(E)=0$ for the resonant energies and there is only the
probability density $\left\vert \psi_{>}(x)\right\vert ^{2}=1/\sqrt{2\pi
}\sigma$ ($x<0$) stemming from an incoming wave and arriving to the $x>d$
area. Thus, the dependence of the constant in space transmitted probability
density (\ref{32}) versus the potential amplitude $U$ will exhibit the
oscillating (at $E_{\bot}>U$) pattern beyond the barrier ($x>d$) with an
amplitude which is greater for the asymmetric barrier ($\Delta\neq0$) as
compared to the symmetric one ($\Delta=0$). The same is true for the
oscillating $x$-dependence of $\left\vert \psi_{>}(x)\right\vert ^{2}$ inside
the potential area ($0<x<d$). \ \ 

The time dependence of the probability density $\left\vert \psi
(x,t)\right\vert ^{2}$ exhibits itself only when there is a sufficient
momentum dispersion, as follows from Eqs. (\ref{27}). On the other hand, a
sufficient momentum dispersion, when $\widetilde{E}_{\bot}\widetilde{\sigma
}^{2}\sim1$, leads to a nonnegligible counterintuitive contribution of the
backward-moving components of the wave packet to $\left\vert \psi
(x,t)\right\vert ^{2}$. The spacetime evolution of the scattering process can
be visualized by numerical evaluation of the probability density $\left\vert
\psi(\widetilde{x},\widetilde{t})\right\vert ^{2}$ (\ref{27}) of finding the
particle in the scaled space-time point ($\widetilde{x},\widetilde{t}$). We
will focus on the influence of the wave packet backward-moving components and
the potential asymmetry parameter $\Delta$ on the particle dynamics. As
mentioned earlier, the asymmetric rectangular potential can model the
potential profile of the magnetic threelayer when it is switched from the
parallel configuration of the magnetic layer (modelled by the symmetric
potential profile with $\Delta=0$) to the antiparallel orientation. For the
case under consideration, when the particle, associated with the Gaussian wave
packet, moves towards the potential (\ref{1}) from the left, one can expect
that the influence of the asymmetry parameter $\Delta$ (defining the height of
the right potential step of (\ref{1})) will be more pronounced if the
contribution of the backward-moving components of the wave packet is essential
(the numerical evaluation confirms this expectation).

To make the dynamics of the wave packet more particle-like, we accept the
condition of the narrow wave packet, $\widetilde{\sigma}<1$, and put
$\widetilde{t}_{0}=0$. For an electron and the potential width $d=10^{-7}cm$
($1nm$), the characteristic energy $E_{d}\sim3\cdot10^{-2}ev$ and the
characteristic time $t_{d}\sim2\cdot10^{-14}s$. In accordance with the
accepted conditions, we will posit $\widetilde{E}_{\bot}=10^{2}$,
$\widetilde{x}_{i}=-10$, and $\widetilde{\sigma}=1/3$ or $\widetilde{\sigma
}=0.1$. We choose $\widetilde{U}=10$ in the case of a potential barrier
(over-barrier transmission), and $\widetilde{U}=-10^{2}$ for a potential well.
We will compare two cases: $\widetilde{\sigma}=1/3$, when the second
inequality (\ref{29}) is satisfied and the backward-moving positive energies
components of the initial wave packet are not essential, and $\widetilde
{\sigma}=0.1$, when their contribution matters. The dimensionless time
interval $\widetilde{t}=0.1\div1.5$ is chosen from a simple estimation for the
average scaled time $t_{i}/t_{d}$ that it takes a particle with the initial
energy $\widetilde{E}_{\bot}=10^{2}$ to reach the potential starting from the
point $\widetilde{x}_{i}=-10$: $t_{i}/t_{d}=\left\vert x_{i}\right\vert
m/\hbar k_{i}t_{d}=\left\vert \widetilde{x}_{i}\right\vert /2\sqrt
{\widetilde{E}_{\bot}}=1/2$.

Figure 1 shows the probability density $\left\vert \psi(\widetilde
{x},\widetilde{t})\right\vert ^{2}$ of finding the particle at $\widetilde
{x}=1$, i.e. on the right-hand side of the barrier (\ref{1}) ($\widetilde
{U}>0$), as a function of $\widetilde{t}$ and $\widetilde{\Delta}$ changing
from $\widetilde{\Delta}=0$ to $\widetilde{\Delta}=\widetilde{E}_{\bot}/2$
when $\widetilde{\sigma}=1/3$. Figure 2 shows the same function for
$\widetilde{\sigma}=0.1$. We see that in the case when the contribution of the
backward-moving components of the wave packet is important ( $\widetilde
{\sigma}=0.1$), the time distribution of finding the particle beyond the
barrier $\left\vert \psi(1,\widetilde{t})\right\vert ^{2}$ for the asymmetric
potential is essentially different from that for the symmetric one: Beginning
from the value of the asymmetry parameter $\widetilde{\Delta}\approx20$, this
distribution becomes more broad and pronouncedly nonmonotonic for
$\widetilde{\Delta}>20$.%

\begin{figure}
[ptb]
\begin{center}
\includegraphics[
height=2.0617in,
width=3.0441in
]%
{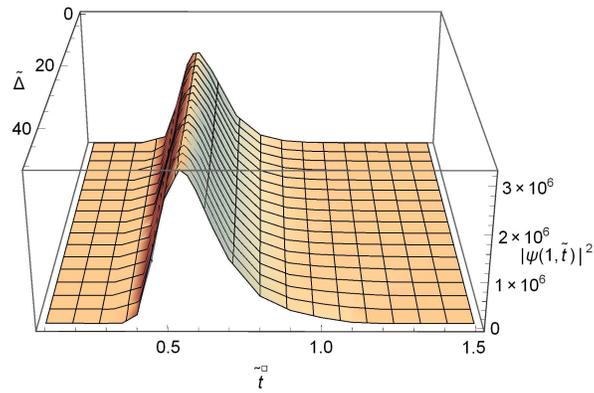}%
\caption{Probability density distribution $\left\vert \psi(1,\widetilde
{t})\right\vert ^{2}$ on the right-hand side of the barrier as a function of
time and asymmetry parameter $\widetilde{\Delta}$ for the narrow energy
distribution of the initial wave packet ($\widetilde{\sigma}=1/3$).}%
\label{Fig1}%
\end{center}
\end{figure}
\begin{figure}
[ptb]
\begin{center}
\includegraphics[
height=2.6204in,
width=3.4221in
]%
{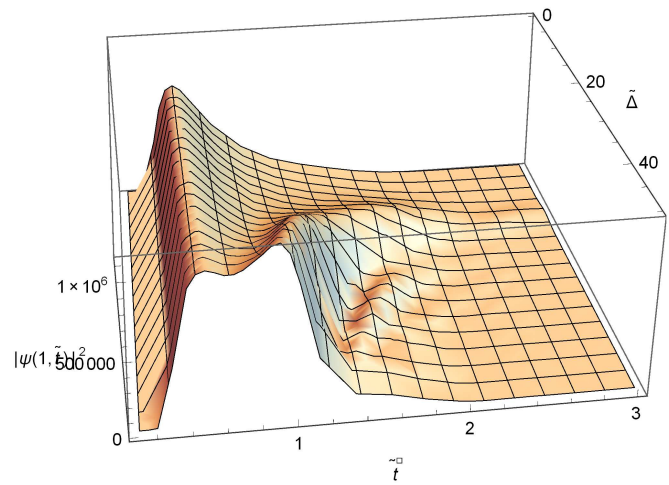}%
\caption{Probability density $\left\vert \psi(1,\widetilde{t})\right\vert
^{2}$ as a function of $\widetilde{t}$ and $\widetilde{\Delta}$ for the broad
energy distribution of the initial wave packet ($\widetilde{\sigma}=0.1$).}%
\label{Fig2}%
\end{center}
\end{figure}
For the case of a potential well with $\widetilde{U}=-10^{2}$, we numerically
evaluated $\left\vert \psi(\widetilde{x},\widetilde{t})\right\vert ^{2}$
inside the well ($\widetilde{x}=0\div1$) for the case of the broad energy
distribution of the initial wave packet ($\widetilde{\sigma}=0.1$) and the
asymmetry parameter $\widetilde{\Delta}=0$ and $\widetilde{\Delta}%
=\widetilde{E}_{\bot}/2=\left\vert \widetilde{U}\right\vert /2$. Figure 3
shows the interference pattern inside the symmetric well which differs
sufficiently from the stationary square cosine type picture, given by Eq.
(\ref{32}) (for $\Delta=0$). It is seen that the amplitude of this pattern
grows with time from zero to the maximum value (reached approximately at
$\widetilde{t}=0.5$) and then again diminishes to zero, thereby showing the
finite time during which a particle exists in the well region before leaving
it either for the region before ($\widetilde{x}<0$) or beyond ($\widetilde
{x}>1$) the well. We also see that the interference pattern of $\left\vert
\psi(\widetilde{x},\widetilde{t})\right\vert ^{2}$ is more structured in space
and time. These changes in the probability density distribution result from
the influence of the backward-moving components of the wave function $\psi
_{<}(\widetilde{x},\widetilde{t})$, which is essential for the considered case
of sufficient energy dispersion ($\widetilde{E}_{\bot}\widetilde{\sigma}%
^{2}=1$). In Fig. 4, we see the influence of the asymmetry parameter
($\widetilde{\Delta}=50$) on that probability density $\left\vert
\psi(\widetilde{x},\widetilde{t})\right\vert ^{2}$ inside the asymmetric well.
The calculated distribution exhibits a very structured and pronouncedly
nonmonotonic interference pattern in space and time compared with that
displayed in Fig. 3.%
\begin{figure}
[ptb]
\begin{center}
\includegraphics[
height=2.3687in,
width=3.0009in
]%
{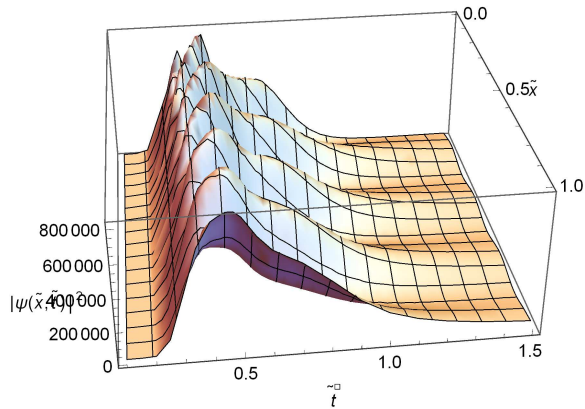}%
\caption{Probability density $\left\vert \psi(\widetilde{x},\widetilde
{t})\right\vert ^{2}$ inside the symmetric well ($\widetilde{\Delta}=0$) for
the broad energy distribution of the initial wave packet ($\widetilde{\sigma
}=0.1$).}%
\label{Fig3}%
\end{center}
\end{figure}
\begin{figure}
[ptb]
\begin{center}
\includegraphics[
height=1.9078in,
width=3.2258in
]%
{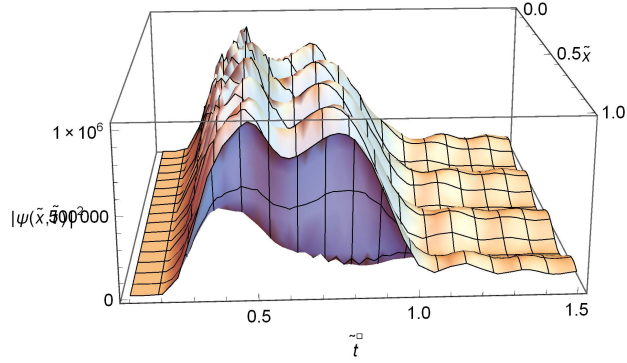}%
\caption{Probability density $\left\vert \psi(\widetilde{x},\widetilde
{t})\right\vert ^{2}$ inside the asymmetric well ($\widetilde{\Delta}=50$) for
$\widetilde{\sigma}=0.1$.}%
\label{Fig4}%
\end{center}
\end{figure}

\section{Dwell time}

For a finite spatial interval, the so-called dwell time, i.e. the average time
spent in this interval by a particle described by the packet $\psi(x,t) $, is
customarily used. The dwell time in the potential region ($0,d$) can be
defined in the three-dimensional case for the potential (\ref{1}) as%
\begin{equation}
\tau(0,d)=\lim_{t-t_{0}\rightarrow\infty}%
{\displaystyle\int\limits_{t_{0}}^{t}}
dt%
{\displaystyle\int\limits_{0}^{d}}
dx\int d\mathbf{\rho}\left\vert \psi(x,\mathbf{\rho,}t)\right\vert ^{2}.
\label{34}%
\end{equation}
Substituting the third and fourth lines of (\ref{19}) into the definition
(\ref{34}), we obtain for the initial wave packet (\ref{22})%

\begin{align}
\tau(0,d)  &  =\tau_{>}(0,d)+\tau_{<}(0,d)+2\operatorname{Re}\tau
_{><}(0,d),\nonumber\\
\tau_{>}(0,d)  &  =%
{\displaystyle\int\limits_{0}^{\infty}}
dE%
{\displaystyle\int\limits_{0}^{d}}
dx\left\vert \frac{\varphi(x,E)}{\sqrt{v_{u}(E)}}\psi_{>}(E)\right\vert
^{2},\tau_{<}(0,d)=%
{\displaystyle\int\limits_{0}^{\infty}}
dE%
{\displaystyle\int\limits_{0}^{d}}
dx\left\vert \frac{\varphi(x,E)}{\sqrt{v_{u}(E)}}\psi_{<}(E)\right\vert
^{2},\nonumber\\
\tau_{><}(0,d)  &  =%
{\displaystyle\int\limits_{0}^{\infty}}
dE%
{\displaystyle\int\limits_{0}^{d}}
dx\left[  \frac{\varphi(x,E)}{\sqrt{v_{u}(E)}}\right]  ^{2}\psi_{>}(E)\psi
_{<}^{\ast}(E),\nonumber\\
\varphi(x,E)  &  =t^{\prime}(E)e^{ik_{u}x}+r^{\prime}(E)e^{-ik_{u}x},
\label{35}%
\end{align}
where the functions $\psi_{>(<)}(E)$ are defined by (\ref{23}) and the
equality (\ref{23a}) is taken into account.

We see that, again, the dwell time is determined by the forward- and
backward-moving components of the initial wave packet as well as their
interference. The entire range of energy ($0\div\infty$) contributes to the
dwell time. It is not difficult to get from (\ref{35}), (\ref{14}) and
(\ref{17}) that the forward- and backward-moving components of the dwell time
are%
\begin{equation}
\tau_{>(<)}(0,d)=%
{\displaystyle\int\limits_{0}^{\infty}}
dE\tau_{>(<)}(E;d). \label{36}%
\end{equation}
The per unit energy interval energy-dependent dwell time $\tau_{>(<)}(E;d)$
caused by the forward- (backward-) moving component of the initial wave packet
$\psi_{>(<)}(E)$ is for $E>\Delta$ ($k_{\Delta}=\sqrt{2m(E-\Delta)/\hbar^{2}}$
is real)
\begin{align}
\tau_{>(<)}(E;d)  &  =t(E;d)\left\vert \psi_{>(<)}(E)\right\vert
^{2},\nonumber\\
t(E;d)  &  =\frac{k}{v_{u}}\frac{2k_{u}d(k_{u}^{2}+k_{\Delta}^{2})-K_{0}%
^{2}\sin(2k_{u}d)}{(k+k_{\Delta})^{2}k_{u}^{2}+k_{0}^{2}K_{0}^{2}\sin
^{2}(k_{u}d)},E>\Delta\geq0, \label{37}%
\end{align}
where $K_{0}^{2}=k_{\Delta}^{2}-k_{u}^{2}=\frac{2m}{\hbar^{2}}(U-\Delta)$,
$k_{0}^{2}=k^{2}-k_{u}^{2}=\frac{2m}{\hbar^{2}}U$. For $E<\Delta$, $\Delta
\geq0$ ($k_{\Delta}=i\overline{k}_{\Delta}$, $\overline{k}_{\Delta}%
=\sqrt{2m(\Delta-E)/\hbar^{2}}$)
\begin{equation}
t(E;d)=\frac{k}{v_{u}}\frac{2k_{u}d(k_{u}^{2}+\overline{k}_{\Delta}%
^{2})-\overline{K}_{0}^{2}\sin(2k_{u}d)+4k_{u}\overline{k}_{\Delta}\sin
^{2}(k_{u}d)}{(k+\overline{k}_{\Delta})^{2}k_{u}^{2}+k_{0}^{2}\overline{K}%
_{0}^{2}\sin^{2}(k_{u}d)+k_{0}^{2}k_{u}\overline{k}_{\Delta}\sin(2k_{u}%
d)},E<\Delta,\Delta>0, \label{37a}%
\end{equation}
where $\overline{K}_{0}^{2}=\overline{k}_{\Delta}^{2}-k_{u}^{2}=\frac
{2m}{\hbar^{2}}(U+\Delta-2E)$. Both Eq. (\ref{37}) and Eq. (\ref{37a}) are
valid for $k_{u}$ real as well as for the imaginary $k_{u}=i\overline{k}_{u}$,
$\overline{k}_{u}=\sqrt{2m(U-E)/\hbar^{2}}$ when $E<U$ (barrier). Note that
$\psi_{>(<)}(E)$ has the dimensionality of the inverse square root of energy
(see (\ref{23})), and thus the expression for $t(E;d)$ has the dimensionality
of time and represents the generalization of the energy-dependent dwell time
obtained earlier by Buttiker \cite{Buttiker 1983} to the case of the
asymmetric rectangular potential (\ref{1}) (if $\Delta=0$, Eq. (\ref{37})
reduces to the Buttiker result).

The per unit energy interval interference dwell time which follows from
(\ref{35}) can be written as%
\begin{align}
\tau_{><}(E;d)  &  =\operatorname{Re}\left\{  \frac{k}{v_{u}}\frac{(k_{u}%
^{2}+k_{\Delta}^{2})\sin(2k_{u}d)+2(k_{u}^{2}-k_{\Delta}^{2})k_{u}%
d-4ik_{u}k_{\Delta}\sin^{2}(k_{u}d)}{\left[  i(kk_{\Delta}+k_{u}^{2}%
)\sin(k_{u}d)-k_{u}(k+k_{\Delta})\cos(k_{u}d)\right]  ^{2}}\psi_{>}(E)\psi
_{<}^{\ast}(E)\right\}  ,\nonumber\\
\operatorname{Re}\tau_{><}(0,d)  &  =%
{\textstyle\int\limits_{0}^{\infty}}
dE\tau_{><}(E;d). \label{38}%
\end{align}
Note that Eq. (\ref{38}) holds for both $k_{u}$ real ($E>U$) and imaginary
$k_{u}=i\overline{k}_{u}$ ($E<U$), as well as for both the real $k_{\Delta}$
($E>\Delta$) and imaginary $k_{\Delta}=i\overline{k}_{\Delta}$ ($E<\Delta$).

We see that the total per unit energy dwell time
\begin{equation}
\tau(E;d)=t(E;d)\left[  \left\vert \psi_{>}(E)\right\vert ^{2}+\left\vert
\psi_{<}(E)\right\vert ^{2}\right]  +2\tau_{><}(E;d),\tau(0,d)=\int
\limits_{0}^{\infty}\tau(E;d)dE \label{39}%
\end{equation}
is generally defined by both the forward- and backward-moving components of
the initial wave packet as well as their interference. For the resonance
energies satisfying the condition $k_{u}d=\pi n$ ($n$ is integer, $k_{u}$ is
real), taking place in the cases of $U<0$ and $U>0$ (when $E>U$), Eqs.
(\ref{37}) -(\ref{39}) reduce, e.g. for $k_{\Delta}$ real ($E\geq\Delta$), to
\begin{equation}
\tau(E;d)=\frac{d}{v}\left\{  \frac{2k^{2}(k_{u}^{2}+k_{\Delta}^{2})}%
{k_{u}^{2}(k+k_{\Delta})^{2}}\left[  \left\vert \psi_{>}(E)\right\vert
^{2}+\left\vert \psi_{<}(E)\right\vert ^{2}\right]  +4\operatorname{Re}%
\frac{k^{2}(k_{u}^{2}-k_{\Delta}^{2})}{k_{u}^{2}(k+k_{\Delta})^{2}}\left[
\psi_{>}(E)\psi_{<}^{\ast}(E)\right]  \right\}  , \label{40}%
\end{equation}
where $d/v(E)$ is the time that it takes for a particle with the energy $E$ to
propagate through the spatial range $d$ in the absence of a potential. Thus,
the expression in the curly brackets in (\ref{40}) shows the difference
between the dwell time in the range of the potential and the "free" dwell time
$d/v(E)$.

From the above it follows that, generally, the dwell time depends on the
energy spectrum of the initial wave packet $\psi_{>(<)}(E)$ and cannot be
realistically defined, e.g., simply by $t(E;d)$ (\ref{37}) or (\ref{37a}).
Further, we will use $\psi_{>(<)}(E)$ (\ref{23}), defined for the Gaussian
initial wave packet, and shift to the dimensionless variables defined in the
previous section. As a result, we obtain from Eqs. (\ref{37}) - (\ref{39})%

\begin{align}
\tau(0,d)  &  =%
{\displaystyle\int\limits_{0}^{\infty}}
d\widetilde{E}\widetilde{\tau}(\widetilde{E};d),\widetilde{\tau}(\widetilde
{E};d)=\widetilde{\tau}_{>}(\widetilde{E};d)+\widetilde{\tau}_{<}%
(\widetilde{E};d)+\widetilde{\tau}_{><}(\widetilde{E};d),\nonumber\\
\widetilde{\tau}_{>}(\widetilde{E};d)+\widetilde{\tau}_{<}(\widetilde{E};d)
&  =\frac{t_{d}\widetilde{\sigma}}{2\sqrt{2\pi}}\frac{1}{\sqrt{\widetilde
{E}-\widetilde{U}}}\frac{2\sqrt{\widetilde{E}-\widetilde{U}}(2\widetilde
{E}-\widetilde{U}-\widetilde{\Delta})-(\widetilde{U}-\widetilde{\Delta}%
)\sin(2\sqrt{\widetilde{E}-\widetilde{U}})}{(\sqrt{\widetilde{E}}%
+\sqrt{\widetilde{E}-\widetilde{\Delta}})^{2}(\widetilde{E}-\widetilde
{U})+\widetilde{U}(\widetilde{U}-\widetilde{\Delta})\sin^{2}(\sqrt
{\widetilde{E}-\widetilde{U}})}\nonumber\\
&  \times\left\{  \exp[-2(\sqrt{\widetilde{E}_{\bot}}-\sqrt{\widetilde{E}%
})^{2}\widetilde{\sigma}^{2}]+\exp[-2(\sqrt{\widetilde{E}_{\bot}}%
+\sqrt{\widetilde{E}})^{2}\widetilde{\sigma}^{2}]\right\}  ,\widetilde
{E}>\widetilde{\Delta}\geq0,\nonumber\\
\widetilde{\tau}_{><}(\widetilde{E};d)  &  =\frac{t_{d}\widetilde{\sigma}%
}{\sqrt{2\pi}}\times\nonumber\\
&  \operatorname{Re}\frac{1}{\sqrt{\widetilde{E}-\widetilde{U}}}%
\frac{(2\widetilde{E}-\widetilde{U}-\widetilde{\Delta})\sin(2\sqrt
{\widetilde{E}-\widetilde{U}})+2\sqrt{\widetilde{E}-\widetilde{U}}%
(\widetilde{\Delta}-\widetilde{U})-4i\sqrt{\widetilde{E}-\widetilde{\Delta}%
}\sqrt{\widetilde{E}-\widetilde{U}}\sin^{2}(\sqrt{\widetilde{E}-\widetilde{U}%
})}{[i(\sqrt{\widetilde{E}}\sqrt{\widetilde{E}-\widetilde{\Delta}}%
+\widetilde{E}-\widetilde{U})\sin(\sqrt{\widetilde{E}-\widetilde{U}}%
)-(\sqrt{\widetilde{E}}+\sqrt{\widetilde{E}-\widetilde{\Delta}})\sqrt
{\widetilde{E}-\widetilde{U}}\cos(\sqrt{\widetilde{E}-\widetilde{U}})]^{2}%
}e^{-2i\sqrt{\widetilde{E}}\widetilde{x}_{i}}\nonumber\\
&  \times\exp[-(\sqrt{\widetilde{E}_{\bot}}-\sqrt{\widetilde{E}}%
)^{2}\widetilde{\sigma}^{2}]\exp[-(\sqrt{\widetilde{E}_{\bot}}+\sqrt
{\widetilde{E}})^{2}\widetilde{\sigma}^{2}], \label{41}%
\end{align}
where the characteristic time $t_{d}/2=\hbar/2E_{d}=\frac{d}{v(E)}%
\sqrt{\widetilde{E}}=\frac{d}{v(E_{\bot})}\sqrt{\widetilde{E}_{\bot}}$, i.e.
it is the time spent in the region of the potential width $d$ by a "free"
particle with the energy $E=E_{d}$ ($\widetilde{E}=1$), and thus
$\widetilde{\tau}(\widetilde{E};d)$ has the dimensionality of time (for
brevity, we do not show Eq. (\ref{37a}) in the dimensionless variables). The
relative contribution of the forward- (backward-) moving components
$\widetilde{\tau}_{>(<)}(\widetilde{E};d)$ and interference term
$\widetilde{\tau}_{><}(\widetilde{E};d)$ to the dwell time $\tau(0,d)$
(\ref{41}) depends on the value of the parameter $\widetilde{E}_{i}%
\widetilde{\sigma}^{2}$. If the second inequality (\ref{29}) is satisfied,
i.e., $\widetilde{E}_{\bot}\widetilde{\sigma}^{2}\gg1$, the contribution of
the backward-moving and interference terms to the dwell time (\ref{41}) is
much smaller than that of the forward-moving term $\widetilde{\tau}%
_{>}(\widetilde{E};d)$, and, therefore, the former terms may be ignored in the
first approximation in the limit given by (\ref{29}). Moreover, the integral
of $\widetilde{\tau}_{>}(\widetilde{E};d)$ over $\widetilde{E}$ can be
asymptotically estimated due to the sharp maximum of the integrand at
$\widetilde{E}=\widetilde{E}_{\bot}$. The result is%
\begin{align}
\tau(0,d)  &  \sim\tau_{>}(0,d)=\frac{d}{v(E_{\bot})}\frac{\widetilde{E}%
_{\bot}}{\sqrt{\widetilde{E}_{\bot}-\widetilde{U}}}\frac{2\sqrt{\widetilde
{E}_{\bot}-\widetilde{U}}(2\widetilde{E}_{\bot}-\widetilde{U}-\widetilde
{\Delta})-(\widetilde{U}-\widetilde{\Delta})\sin(2\sqrt{\widetilde{E}_{\bot
}-\widetilde{U}})}{(\sqrt{\widetilde{E}_{\bot}}+\sqrt{\widetilde{E}_{\bot
}-\widetilde{\Delta}})^{2}(\widetilde{E}_{\bot}-\widetilde{U})+\widetilde
{U}(\widetilde{U}-\widetilde{\Delta})\sin^{2}(\sqrt{\widetilde{E}_{\bot
}-\widetilde{U}})},\nonumber\\
\widetilde{E}_{\bot}  &  >\widetilde{\Delta}\geq0, \label{42}%
\end{align}
which coincides with Eq. (\ref{37}) for $t(E_{\bot};d)$ written in the
dimensionless variables. It should be stressed that this result represents
only the first term of the asymptotic expansion of $%
{\displaystyle\int\limits_{0}^{\infty}}
d\widetilde{E}\tau_{>}(\widetilde{E};d)$ with a small value of the parameter
$1/\widetilde{E}_{\bot}\widetilde{\sigma}^{2}$, i.e. for an initial wave
packet characterized by an extra narrow momentum distribution.

For the resonance energies satisfying the condition $k_{u}d=\pi n$ ($n$ is
integer, $n\neq0$, $k_{u}$ is real), which reads in the dimensionless
variables as $\widetilde{E}_{\bot}-\widetilde{U}=\pi^{2}n^{2}$, the relative
to the "free" dwell time $d/v(E_{\bot})$ expression (\ref{42}) reduces to%
\begin{equation}
\tau_{>}^{r}(0,d)v(E_{\bot})/d=\frac{2\widetilde{E}_{\bot}(2\widetilde
{E}_{\bot}-\widetilde{U}-\widetilde{\Delta})}{(\sqrt{\widetilde{E}_{\bot}%
}+\sqrt{\widetilde{E}_{\bot}-\widetilde{\Delta}})^{2}(\widetilde{E}_{\bot
}-\widetilde{U})}=\frac{2\widetilde{E}_{\bot}(\widetilde{E}_{\bot}%
-\widetilde{\Delta}+\pi^{2}n^{2})}{(\sqrt{\widetilde{E}_{\bot}}+\sqrt
{\widetilde{E}_{\bot}-\widetilde{\Delta}})^{2}\pi^{2}n^{2}}. \label{43}%
\end{equation}
At $\widetilde{U}<0$ (dwell), the inequality $\left\vert \widetilde
{U}\right\vert =\pi^{2}n^{2}-\widetilde{E}_{\bot}>0$ should be satisfied ($n$
is bottom-limited), and when $\pi^{2}n^{2}\gg>\widetilde{E}_{\bot}%
>\widetilde{E}_{\bot}-\widetilde{\Delta}\geq0$, which is the case for large
enough $\left\vert \widetilde{U}\right\vert $, the asymptotic relative
resonant dwell time (\ref{43}) approaches $2\widetilde{E}_{\bot}%
/(\sqrt{\widetilde{E}_{\bot}}+\sqrt{\widetilde{E}_{\bot}-\widetilde{\Delta}%
})^{2}$. This value is greater than $1/2$, to which the values of the high
order resonances of the dwell time reduce for a symmetric potential
($\widetilde{\Delta}=0$). Thus, the greater the asymmetry parameter $\Delta$,
the greater the amplitudes of the dwell time resonances. For $\widetilde{U}>0$
(barrier), the condition $\widetilde{U}=\widetilde{E}_{\bot}-\pi^{2}n^{2}>0 $
should hold ($\widetilde{E}_{\bot}>\widetilde{U}$, $n$ is restricted to the
small values defined by $\widetilde{E}_{\bot}$), and at $\pi^{2}n^{2}%
\ll\widetilde{E}_{\bot}-\widetilde{\Delta}<\widetilde{E}_{\bot}$ the dwell
time (\ref{43}) behaves as $2\widetilde{E}_{\bot}(\widetilde{E}_{\bot
}-\widetilde{\Delta})/(\sqrt{\widetilde{E}_{\bot}}+\sqrt{\widetilde{E}_{\bot
}-\widetilde{\Delta}})^{2}\pi^{2}n^{2}$.

If $\widetilde{E}_{\bot}-\widetilde{U}\rightarrow0$ (reverse points in
classical physics), which can happen only at $\widetilde{U}>0$, the asymptotic
relative dwell time $\tau_{>}(0,d)v(E_{\bot})/d$ (\ref{42}) reduces to%
\begin{equation}
\tau_{>}(0,d)v(E_{\bot})/d=\frac{4\widetilde{E}_{\bot}}{(\sqrt{\widetilde
{E}_{\bot}}+\sqrt{\widetilde{E}_{\bot}-\widetilde{\Delta}})^{2}+\widetilde
{U}(\widetilde{U}-\widetilde{\Delta})}>\frac{4\widetilde{E}_{\bot}%
}{4\widetilde{E}_{\bot}+\widetilde{U}^{2}}<1,\widetilde{E}_{\bot}%
-\widetilde{U}\rightarrow0, \label{44}%
\end{equation}
\ where the value $4\widetilde{E}_{\bot}/(4\widetilde{E}_{\bot}+\widetilde
{U}^{2})$ corresponds to a symmetric potential, i.e., the dwell time
(\ref{44}) in the asymmetric case is larger. In particular, at $\widetilde
{U}=0$ this dwell time for $\Delta>0$ is larger than the "free" dwell time
$d/v(E_{\bot})$ in the absence of a potential ($\widetilde{U}=0$, $\Delta=0$).

It is interesting to plot the dependence of the relative dwell time (\ref{42})
on the amplitude of the potential $\widetilde{U}$, which changes from negative
values (the well) to positive ones (the barrier), for a symmetric
($\widetilde{\Delta}=0$) and an asymmetric potential. Formula (\ref{42}) is
valid for both the $\widetilde{U}<0$ and $\widetilde{U}>0$ cases, and in the
latter case, when $\widetilde{E}_{\bot}-\widetilde{U}<0$, Eq. (\ref{42})
transfers to%
\begin{equation}
\tau_{>}(0,d)v(E_{\bot})/d=\frac{\widetilde{E}_{\bot}}{\sqrt{\widetilde
{U}-\widetilde{E}_{\bot}}}\frac{2\sqrt{\widetilde{U}-\widetilde{E}_{\bot}%
}(\widetilde{U}+\widetilde{\Delta}-2\widetilde{E}_{\bot})+(\widetilde
{U}-\widetilde{\Delta})\sinh(2\sqrt{\widetilde{U}-\widetilde{E}_{\bot}}%
)}{(\sqrt{\widetilde{E}_{\bot}}+\sqrt{\widetilde{E}_{\bot}-\widetilde{\Delta}%
})^{2}(\widetilde{U}-\widetilde{E}_{\bot})+\widetilde{U}(\widetilde
{U}-\widetilde{\Delta})\sinh^{2}(\sqrt{\widetilde{U}-\widetilde{E}_{\bot}})}.
\label{45}%
\end{equation}
%

\begin{figure}
[ptb]
\begin{center}
\includegraphics[
height=2.3964in,
width=3.659in
]%
{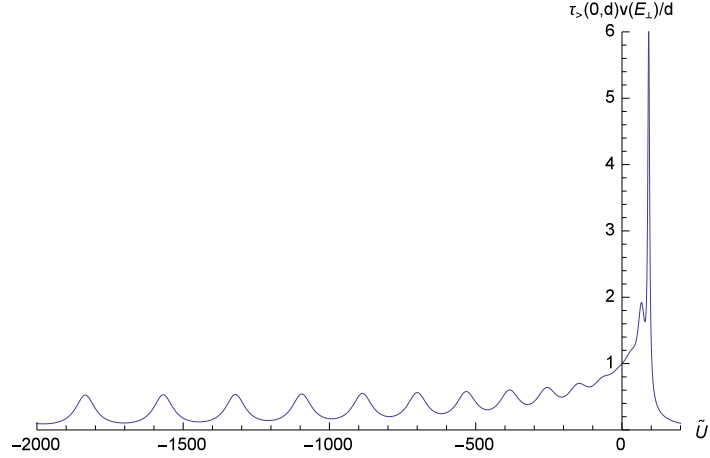}%
\caption{Dependence of the asymptotic dwell time (\ref{42}) on the
well/barrier symmetric potential ($\widetilde{\Delta}=0$).}%
\label{Fig5}%
\end{center}
\end{figure}
\begin{figure}
[ptb]
\begin{center}
\includegraphics[
height=2.3964in,
width=3.659in
]%
{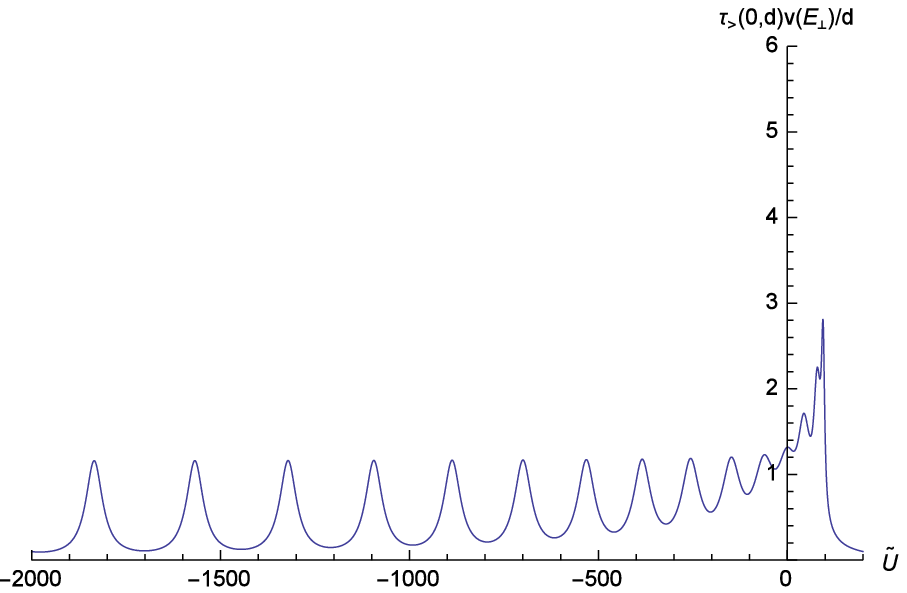}%
\caption{Dependence of the asymptotic dwell time (\ref{42}) on the
well/barrier asymmetric potential ($\widetilde{\Delta}=90$).}%
\label{Fig6}%
\end{center}
\end{figure}
Fig. \ref{Fig5} shows the $\widetilde{U}$ - dependence of $\tau_{>}%
(0,d)v(E_{\bot})/d$ (\ref{42}) for a symmetric potential ($\widetilde{\Delta
}=0$) in the broad range of $\widetilde{U}=-2\times10^{3}\div2\times10^{2}$
and $\widetilde{E}_{\bot}=10^{2}$. One can see the series of resonances at
$\widetilde{U}<0$, the amplitudes of which approach $1/2$ for big enough $n$,
$n^{2}\gg\widetilde{E}_{\bot}/\pi^{2}$, while at $\widetilde{U}>0$ there is a
limited series of resonances with $n^{2}<\widetilde{E}_{\bot}/\pi^{2}$ (for
$\widetilde{E}_{\bot}>\widetilde{U}$) with the larger amplitudes because a
particle moves more slowly in the presence of a potential barrier than in the
region of a potential well. Fig. \ref{Fig6} shows the same $\widetilde{U}$ -
dependence of $\tau_{>}(0,d)v(E_{\bot})/d$ (\ref{42}) for $\widetilde{\Delta
}=90$. We see an essential increase of the resonances' amplitudes inside the
well and other details which display the influence of the potential asymmetry
on the dwell time in correspondence with the analysis given above.

\section{Summary}

We have applied the MST to the calculation of the propagator which exactly
resolves the time-dependent Schr\"{o}dinger equation for a particle in the
presence of a one-dimensional rectangular asymmetric well/barrier potential
(\ref{1}). This approach, based on the obtained effective potentials
(\ref{7}), (\ref{8}), which are responsible for reflection from and
transmission through the potential steps, is alternative to the matching
procedure conventionally used for solving the stationary Schr\"{o}dinger
equation. The advantages of this MST approach are: A natural picture of the
considered processes in terms of a particle scattering at the potential jumps
(in contrast to the traditional wave point of view); The time-dependent
picture of the quantum effects of particle reflection from a potential well
and particle transmission through a potential barrier; The natural
decomposition of the Schr\"{o}dinger equation solution into the sum of the
forward- and backward-moving terms (with no use of the evanescent states
\cite{Muga Sources 2001}), which takes into account that the initial wave
packet, confined to a restricted spatial area and representing a particle
moving towards a potential, contains both the positive and negative momentum
components. Aside from being related to the fundamental issues of quantum
mechanics, the obtained results can be also important for the kinetic theory
of nanostructures, where the considered rectangular potential (\ref{1}) is
often used to model the potential profile in the magnetic nanostructures
utilized, e.g., in spintronics devices.

The obtained probability density $\left\vert \psi(x,t)\right\vert ^{2}$ of
finding a particle in the space-time point $(x,t)$, when it initially was
located in some spatial region and moved in some direction, is generally
defined by the probability density corresponding to the wave component moving
in this direction $\left\vert \psi_{>}(x,t)\right\vert ^{2}$ as well as by the
probability densities related to the backward-moving component $\left\vert
\psi_{<}(x,t)\right\vert ^{2}$ and the interference of both
$2\operatorname{Re}[\psi_{>}(x,t)\psi_{<}^{\ast}(x,t)]$. For the case of the
initial Gaussian wave packet, we have shown that the contribution of the
backward-moving component to the probability density $\left\vert
\psi(x,t)\right\vert ^{2}$ is small when the initial packet is characterized
by a narrow energy (momentum) distribution, which is characteristic of the
quasi-classical approximation for a transport phenomenon. We calculated, in
this case, the asymptotic time-independent values of $\left\vert \psi
_{>}(x)\right\vert ^{2}$ in the different spatial regions relative to the
potential area. This situation (extra narrow energy distribution) actually
corresponds to the stationary case with no energy dispersion. Thus, the
transmission through and reflection from the potential well/barrier can be
described as a function of time only when the momentum (energy) dispersion of
the initial wave packet is significant (accordingly, the wave packet spatial
localization is narrow). But in this case, the counterintuitive
(non-classical) contribution of the backward-moving components of the wave
packet should be accounted for. This rather paradoxical quantum mechanical
result reveals itself in the problems connected to measuring time in quantum
mechanical effects.

Using the exact result for $\left\vert \psi(x,t)\right\vert ^{2}$, we have
numerically plotted the time distribution of finding the particle beyond the
barrier ($U>0$), $\left\vert \psi(1,t)\right\vert ^{2}$ and found that, when
the contribution of the backward-moving wave packet components is important
(broad wave packet energy distribution), the influence of the potential
asymmetry can be essential (Figs. 1,2). Plotting $\left\vert \psi
(x,t)\right\vert ^{2}$ in the well ($U<0$) region, we showed that the
backward-moving components of the wave packet fundamentally change the
probability density, when the initial wave packet is broad enough in the
energy (momentum) space, and the asymmetry of the potential well adds more to
the structure of this spacetime distribution (Figs. 3,4).

The obtained solution is applied to the calculation of the particle time dwell
time within the potential area. Again, the forward- and backward-moving
components of the obtained exact wave function contribute to the particle
dwell time. For a narrow momentum distribution of the initial wave packet, the
analytical asymptotic value of the main (in this case) term contributing to
the dwell time in the potential region, caused by the forward-moving
probability density $\left\vert \psi_{>}(x,t)\right\vert ^{2}$, was obtained
and plotted as a function of the potential amplitude $U$ changing from the
negative (well) to the positive (barrier) values. The series resonances
displayed in Figs. 5,6 show the essential influence of the potential asymmetry
on the particle dwell time. These results generalize the known Buttiker
results \cite{Buttiker 1983} for the dwell time.

\end{document}